\documentclass[aps,prb,10pt,twocolumn,reprint,superscriptaddress,floatfix,amsmath,amssymb]{revtex4-2}

\usepackage{graphicx}% Include figure files
\graphicspath{{figures/}}

\usepackage{dcolumn}% Align table columns on decimal point
\usepackage{bm}% bold math
\usepackage{hyperref}% add hypertext capabilities
\usepackage{mhchem}% Chemistry notations
\usepackage{color}
\usepackage{siunitx}
\usepackage{algorithm}
\usepackage{algpseudocode}
\usepackage[export]{adjustbox}
\newcommand{\Tr}{\operatorname{Tr}}%
\newcommand{\eye}{\mathbf{1}}%
\newcommand{\DD}{\mathcal{D}}%
\newcommand{\CC}{\mathcal{C}}%
\newcommand{\bigO}{\mathcal{O}}%
\newcommand{\VV}{\mathcal{V}}%
\newcommand{\Ss}{\mathcal{S}}%
\newcommand{\TT}{\mathcal{T}}%
\newcommand{\ee}{\mathrm{e}}%
\newcommand{\dd}{\mathrm{d}}%
\newcommand{\ii}{\mathrm{i}}%
\newcommand{\cbar}{\bar{c}}%
\newcommand{\cop}{\hat{c}}%
\newcommand{\cdag}{\hat{c}^{\dagger}}%
\newcommand{\sgn}{\operatorname{sgn}}%
\newcommand{\adj}{\operatorname{adj}}%
\newcommand{\Gmat}{\mathbf{G}}%
\newcommand{\Amat}{\mathbf{A}}%
\newcommand{\kmax}{k_{\max}}%
\newcommand{\iv}{\ii\nu}
\newcommand{\shift}{{\bm{\alpha}}}

\begin{document}

\title{Diagrammatic Monte Carlo Method for Impurity Models with General Interactions and Hybridizations}

\def\umphys{%
Department of Physics, University of Michigan,
Ann Arbor, MI 48109, USA
}%
\def\tuwien{%
Institute of Solid State Physics, Vienna University of Technology,
A-1040 Vienna, Austria
}%

\author{Jia Li}
\affiliation{\umphys}

\author{Markus Wallerberger}
\affiliation{\umphys}
\affiliation{\tuwien}

\author{Emanuel Gull}
\affiliation{\umphys}

\date{\today}

\begin{abstract}
We present a diagrammatic Monte Carlo method for quantum impurity problems with general interactions and general hybridization functions. Our method uses a recursive determinant scheme to sample diagrams for the scattering amplitude. Unlike in other methods for general impurity problems, an approximation of the continuous hybridization function by a finite number of bath states is not needed, and accessing low temperature does not incur an exponential cost.
We test the method for the example of molecular systems, where we systematically vary temperature, interatomic distance, and basis set size.
We further apply the method to an impurity problem generated by a self-energy embedding calculation of correlated antiferromagnetic \ce{NiO}.
We find that the method is ideal for quantum impurity problems with a large number of orbitals but only moderate correlations.
\end{abstract}

\maketitle

\section{Introduction}

Quantum impurity models, originally introduced to describe magnetic impurities such as iron or copper atoms with partially filled d-shells in a non-magnetic host material~\cite{Anderson61}, have since found applications in nanoscience as representations of quantum dots and molecular conductors~\cite{Hanson07}, and in surface science to understand the adsorption of atoms on surfaces~\cite{Brako81,Langreth91}.
In addition, they form the central part of embedding theories such as the dynamical mean field theory (DMFT)~\cite{Georges1996,Kotliar2006} and its variants~\cite{Lichtenstein2000,Kotliar2001,Hettler2000,Maier2005,Anisimov1997,Lichtenstein1998,Held2006,Sun2002,Biermann2003,Biermann2005,Boehnke2016,Choi2016,Lee2017}, as well as the self-energy embedding theory (SEET)~\cite{Kananenka2015,Lan2015,NguyenLan2016}, where they describe the behavior of a few `strongly correlated' orbitals embedded into a weakly correlated or non-interacting background of other orbitals. These methods promise a systematic route for the simulation of strongly correlated quantum many-body problems~\cite{Zgid2017}.

While the original formulation of a quantum impurity model~\cite{Anderson61} only describes a single correlated orbital coupled to a non-interacting environment, in general the impurities occurring in the context of surface science and embedding theories contain many orbitals with general four-fermion interactions and few symmetries \cite{Mazurenko10}.
The time-dependent hybridization function describing the hopping between the impurity and its environment is typically such that it cannot be diagonalized for all frequencies at once.

Solving quantum impurity problems, i.e. obtaining the impurity Green's function given an impurity Hamiltonian and a hybridization function, requires the use of numerical methods.
 A wide range of such methods exist. Hamiltonian-based methods, such as exact diagonalization~\cite{Caffarel1994,Capone2004,Koch2008,Liebsch2009,Senechal2010} and its variants~\cite{Lu14}, configuration-interactions~\cite{Zgid2012}, or coupled cluster theory~\cite{Shee2019,Zhu2019}, solve the impurity problem by mapping the impurity problem onto a system with a local Hamiltonian and a finite number of auxiliary `bath' states chosen to fit the time-dependent hybridization function. The methods  are limited to a relatively small set of strongly interacting sites or break down at moderate correlation strength. The bath fitting, which typically approximates a continuous bath dispersion by a non-linear fit to a small number of delta-function peaks, introduces additional approximations \cite{Koch2008,Senechal2010}.
Numerical renormalization techniques~\cite{Weichselbaum2007,Bulla2008} overcome this issue by providing an almost continuous bath density of states but are in turn limited to a few orbitals in highly symmetrical situations.

A complementary approach is given by Monte Carlo techniques such as the continuous-time quantum Monte Carlo methods~\cite{Gull2011}. These methods are based on a stochastic sampling of the terms in a diagrammatic expansion of the partition function. For particle-hole symmetric systems with on-site density-density interactions, interaction expansion methods~\cite{Rubtsov2005,Gull2008,Gull11} can solve systems with hundreds of strongly correlated orbitals~\cite{LeBlanc15}. Away from particle hole symmetry and at low temperature, they are typically limited to around eight orbitals, and their naive adaptation to general four-fermion operator terms suffers from a severe sign problem \cite{Gorelov09}. In contrast, a partition function expansion in the hybridization~\cite{Werner2006a,Werner2006b,Haule2007} is able to work with general local Hamiltonians of up to five orbitals, but is similarly restricted to diagonal hybridization functions. A reformulation~\cite{Eidelstein19} in terms of `inchworm' diagrams~\cite{Cohen2015} overcomes the restriction of diagonal hybridizations, but so far remains limited to impurities with up to three orbitals.

There is therefore a need for impurity solver methods that can treat the problems of embedding theory and surface science, where several orbitals with general interactions and hybridizations occur.
Diagrammatic Monte Carlo methods~\cite{Prokofev1998b,Prokofev2008a,Prokofev2008b,VanHoucke2010,Chen2019}, which expand physical observables rather than partition functions, along with efficient ways of evaluating the resulting diagrammatic series via the connected determinant (CDet) approach~\cite{Rossi2017b,Rossi2018,Boag2018,Rossi2020,Moutenet2018,Simkovic2019}, are promising. While these methods suffer from other limitations, including divergences of the series in the strong correlation regime, they do not require to approximate the hybridization function by a fit, and are not based on a diagonalization of the local Hamiltonian.

In this paper, we show a formulation of the diagrammatic Monte Carlo method for impurity problems with general interactions and hybridizations based on the CDet framework. We test the method on the example of molecular systems, for which a broad range of very mature Hamiltonian methods exist. From the point of view of the algorithmic formulation, the molecular systems exhibit the full complexity of general impurity problems. The only difference between molecules and quantum impurities is that the latter are formulated with a time-dependent hybridization, rather than an instantaneous hopping. This hybridization function modifies the bare propagator but otherwise leaves the system and our algorithmic approach invariant. Applications to molecular systems therefore form an ideal testbed for impurity solver methods of this type.
We complete our benchmark by applying the impurity solver to an impurity generated by  a self-energy embedding calculation of antiferromagnetic solid \ce{NiO}~\cite{Iskakov2020}.

We carefully analyze the convergence behavior of the diagrammatic expansion and the computational cost of the method as a function of varying temperature, basis sets, intermolecular distance, and system size. We emphasize that we do not intend to present our method as a viable method for quantum chemistry systems without retardation effects. Rather, we exploit the rigorous and controlled framework of molecular simulations to generate a series of test cases that illustrate various parameter regimes in quantum impurities.

This paper will proceed as follows. In Sec.~\ref{sec:method} we introduce the computational problem, the diagrammatic formulation, and the algorithmic description. In Sec.~\ref{sec:results} we present applications to molecular systems
and benchmark results for quantum impurities.
Finally, Sec.~\ref{sec:conclusions} presents conclusions. Appendices \ref{app:ham} through \ref{app:adjugate} present technical details useful for implementing our algorithm and reproducing our results.

\section{Method}\label{sec:method}

\subsection{Partition function expansion}

We describe molecular electrons using the following Hamiltonian:
\begin{equation}
\hat{H} =
\underbrace{\sum_{ab} h_{ab}\cdag_a \cop_b}_{\hat{H}_0} +
\underbrace{\frac{1}{4} \sum_{abcd} U_{abcd} \cdag_a \cdag_c \cop_d \cop_b}_{\hat{H}_V},
    \label{eq:H}
\end{equation}
where $a,b,c,d$ denote spin-orbitals, $1,\ldots,N$. We employ second quantization:
$\cop_a$ and $\cdag_a$ annihilates
and creates, respectively, an electron in the spin-orbital $a$.
The non-interacting term $\hat{H}_0$ is parametrized by the one-electron
integrals $h_{ab} = [a|h|b]$, whereas the interacting term $\hat{H}_V$ is parametrized
by the antisymmetrized two-electron integrals $U_{abcd} = [ab|cd]-[ad|cb]$.
We note that explicit antisymmetrization, $U_{abcd} = -U_{adcb} = U_{cdab}$,
avoids ambiguities in the diagrammatic expansions below~\cite{Motta2017}.
We orthonormalize the basis, $\{\cdag_a, \cop_b\} = \delta_{ab}$, as
we empirically found this to improve the error bars in the subsequent Monte
Carlo procedure.  For completeness, we compiled the explicit expressions for $h$ and $U$
in Appendix~\ref{app:ham}.

% \paragraph{definition of partition function and action}
As we are going to perform series expansions later,
it is convenient to introduce an expansion parameter $\xi$ into the Hamiltonian:
\begin{equation}
   \hat{H}_\xi = \hat{H}_0 + \xi \hat{H}_V.
   \label{eq:Hxi}
\end{equation}
The non-interacting case is given by $\hat{H}_{\xi=0}$, whereas $\hat{H}_{\xi=1}$ recovers
the full Hamiltonian~(\ref{eq:H}).

We are primarily interested in calculating finite temperature observables such as
energies, densities, as well as the spectral function and other
electronic correlation functions.  We start with the grand-canonical
partition function:
\begin{equation}
   Z_\xi = \Tr\exp[-\beta(\hat{H}_\xi - \mu \hat{N})],
   \label{eq:Z}
\end{equation}
where $\beta = 1/T$ is the inverse temperature, $\mu$ denotes
the chemical potential, and $\hat{N}=\sum_{a}\cdag_a \cop_a$ is the density operator.
Expanding Eq.~(\ref{eq:Z}) about $\xi=0$ within the interaction picture~\cite{Abrikosov1965}
yields the Dyson series:
\begin{equation}
  Z_\xi = Z_0 \sum_{k=0}^\infty \frac{(-\xi)^k}{k!} \int_0^\beta
\dd^k\tau \langle \hat{H}_V(\tau_1) \cdots \hat{H}_V(\tau_k) \rangle_0,
  \label{eq:dysonseries}
\end{equation}
where $Z_0 := \Tr\exp[-\beta(\hat{H}_0 - \mu \hat{N})]$ is the non-interacting partition function, $\tau$ denotes imaginary (Euclidean) time, $\langle\cdot\rangle_0$
denotes the non-interacting expectation value:
\begin{equation}
  \langle \hat{X}_1 \ldots \hat{X}_k \rangle_0 := \frac 1{Z_0}
  \Tr[\ee^{-\beta (\hat{H}_0 - \mu \hat{N})} \TT(\hat{X}_1 \ldots \hat{X}_k)],
\end{equation}
and $\TT$ indicates path ordering in imaginary time.  We note that for
molecules, both $\hat{H}_0$ and $\hat{H}_V$ are bounded and thus
away from zero temperature, the series expansion for the
partition function (\ref{eq:dysonseries}) is
absolutely convergent for all $\xi$ .

Inserting Eq.~(\ref{eq:H}) into Eq.~(\ref{eq:dysonseries}) yields:
\begin{equation}
\begin{split}
  \frac{Z_\xi}{Z_0} &= \sum_{k=0}^\infty \frac{(-\xi)^k}{k!}
  \sum_{a_1b_1c_1d_1}\!\cdots\!\sum_{a_kb_kc_kd_k} \int_0^\beta \dd\tau_1\cdots\int_0^\beta \dd\tau_k\\
   &\ \times \bigg(\frac{U_{a_1b_1c_1d_1}}{4}\bigg) \cdots \bigg(\frac{U_{a_kb_kc_kd_k}}{4}\bigg) \big\langle \cdag_{a_1}\!(\tau_1) \cdag_{c_1}\!(\tau_1) \\
   &\ \times \cop_{d_1}\!(\tau_1) \cop_{b_1}\!(\tau_1)
   \ \cdots\ \cdag_{a_k}\!(\tau_k) \cdag_{c_k}\!(\tau_k) \cop_{d_k}\!(\tau_k) \cop_{b_k}\!(\tau_k) \rangle_0 .
\end{split}
\label{eq:Zexp1}
\end{equation}

In order to simplify our notation we combine four spin-orbitals $a,b,c,d$ and
an imaginary time $\tau$ into a single ``vertex''
$v = (a_v,b_v,c_v,d_v,\tau_v)$.  We also introduce the following shorthands:
\begin{subequations}
\begin{align}
  \int \dd^k\VV &:= \frac{1}{k!} \prod_{v\in\VV} \sum_{a_vb_vc_vd_v} \int_0^\beta \dd\tau_v \label{eq:intdv} \\
%  g^{-1}(x,y) &:= (-\partial_{\tau_{x}} + \mu)\delta_{a_xa_y} - h_{a_xa_y} \\
%  U(v) &:= U_{a_vb_{v}c_vd_{v}} \\
  D(\VV) &:= \prod_{v\in\VV} \bigg(\!-\frac{U_{a_vb_{v}c_vd_{v}}}{4}\bigg) \nonumber \\
   &\quad\times \left\langle \prod_{v\in\VV}
    \cdag_{a_v}\!(\tau_v) \cdag_{c_v}\!(\tau_v) \cop_{d_v}\!(\tau_v) \cop_{b_v}\!(\tau_v)
  \right\rangle_{\!\!\!0}. \label{eq:DVV}
\end{align}
\end{subequations}
Eq.~(\ref{eq:DVV}) emphasizes the fact that expectation value
in Eq.~(\ref{eq:Zexp1}) corresponds to the sum over all disconnected and
connected Feynman diagrams with vertices $\VV=(v_1,\ldots,v_k)$, while Eq.~(\ref{eq:intdv})
just corresponds to the sum over all internal degrees of freedom of the
diagrams. With these substitutions, Eq.~(\ref{eq:Zexp1}) simplifies to:
\begin{equation}
  \frac{Z_\xi}{Z_0} = \sum_{k=0}^\infty \xi^k \int \dd^k\VV\ D(\VV).
  \label{eq:Zexp}
\end{equation}

To evaluate Eq.~(\ref{eq:Zexp}), we first introduce the
non-interacting Green's function:
\begin{equation}
  g_{ba}(\tau) = -\langle \cop_b(\tau) \cdag_a(0) \rangle_0
  = [(-\partial_\tau + \mu)\eye - h]^{-1}_{ba}.
  \label{eq:g}
\end{equation}
Given a diagram $\VV = (v_1, \ldots, v_k)$ with $v_i = (a_i, b_i, c_i, d_i, \tau_i)$,
we can use Wick's theorem to write Eq.~(\ref{eq:DVV}) as:
\begin{equation}
  D(\VV) = \prod_{i=1}^{k} \bigg(\!-\frac{U_{a_ib_ic_id_i}}{4}\bigg) \det\Gmat(\VV),
\label{eq:ddqmc}
\end{equation}
where $\Gmat$ is a $2k\times 2k$ matrix in which the
rows (columns) correspond to the $2k$ annihilation (creation) operators.
Introducing the column and row indices $\alpha,\beta,\ldots$ such that
\begin{equation}
\begin{split}
    &\{a_\alpha\} := \{a_1, c_1, a_2, c_2, \ldots, a_k, c_k\},\\
    &\{b_\beta\} := \{b_1, d_1, b_2, d_2, \ldots, b_k, d_k\},\\
    &\{\tau_\alpha\} = \{\tau_\beta\} := \{\tau_1, \tau_1, \tau_2, \tau_2, \ldots, \tau_k, \tau_k\},
\end{split}
\label{eq:colrowidx}
\end{equation}
we define the matrix elements
\begin{align}
    [\Gmat(\VV)]_{\beta\alpha} &:= -\langle \cop_{b_\beta}(\tau_\beta) \cdag_{a_\alpha}(\tau_\alpha) \rangle_0 \nonumber\\
    &= g_{b_\beta a_\alpha}(\tau_\beta-\tau_\alpha + 0^-).
    \label{eq:Gmatelem}
\end{align}
The full matrix can be written
in a block form as
\begin{equation}
    \Gmat(\VV) := \left[\begin{matrix}
        \bm{g}_{11}     & \bm{g}_{12}   & \cdots    & \bm{g}_{1n}\\
        \bm{g}_{21}     & \bm{g}_{22}   & \cdots    & \bm{g}_{2n}\\
        \vdots          & \vdots        & \ddots    & \vdots\\
        \bm{g}_{n1}     & \bm{g}_{n2}   & \cdots    & \bm{g}_{nn}
    \end{matrix}\right],
    \label{eq:Gmat}
\end{equation}
where each $2\times 2$ block is given by
\begin{equation}
    \bm{g}_{ij} := \left[\begin{matrix}
        g_{b_ia_j}(\tau_i - \tau_j + 0^-) & g_{d_ia_j}(\tau_i - \tau_j + 0^-) \\
        g_{b_ic_j}(\tau_i - \tau_j + 0^-) & g_{d_ic_j}(\tau_i - \tau_j + 0^-)
    \end{matrix}\right].
    \label{eq:Gmat-block}
\end{equation}
% \begin{equation}
% \begin{split}
%   &\Gmat(\VV):= \\
%   &\left[\begin{matrix}
% g_{a_1b_1}(0^+)       & \cdots & g_{a_1b_k}(\tau_{n1}) & g_{a_1d_1}(0^+)       & \cdots & g_{a_1d_k}(\tau_{n1}) \\
% \vdots                & \ddots & \vdots                & \vdots                & \ddots & \vdots                \\
% g_{a_kb_1}(\tau_{1n}) & \cdots & g_{a_kb_k}(0^+)       & g_{a_kd_1}(\tau_{1n}) & \cdots & g_{a_kd_k}(0^+)       \\
% g_{c_1b_1}(0^-)       & \cdots & g_{c_1b_k}(\tau_{n1}) & g_{c_1d_1}(0^+)       & \cdots & g_{c_1d_k}(\tau_{n1}) \\
% \vdots                & \ddots & \vdots                & \vdots                & \ddots & \vdots                \\
% g_{c_kb_1}(\tau_{1n}) & \cdots & g_{c_kb_k}(0^-)       & g_{c_kd_1}(\tau_{1n}) & \cdots & g_{c_kd_k}(0^+)       \\
%    \end{matrix}\right]
% \end{split}
% \label{eq:Gmat}
% \end{equation}
% and $\tau_{ij} := \tau_j - \tau_i$.

Eqs.~(\ref{eq:Zexp}) and (\ref{eq:ddqmc}) serve as the basis of interaction expansion continuous-time quantum Monte
Carlo (CT-QMC): one generates random configurations ($v_1\ldots v_k$) and evaluates
the corresponding weight by computing the determinant~\cite{Rubtsov2005,Gull2011,Gorelov09}.

\subsection{Free energy expansion}

While the partition function expansion can be efficiently computed as
determinants (with scaling $\bigO(k^3)$) and
the series is guaranteed to converge, it is also plagued by the negative
sign problem, which is expected to worsen exponentially as the system size is increased
or the temperature reduced.
The sign problem is typically manageable in Hubbard model calculations
up to moderate correlations and system size,
where it only stems from negative determinant contributions. In contrast,
the sign problem is particularly severe in molecules and surface science quantum impurity problems \cite{Gorelov09},
where both Coulomb interaction terms and determinants generate negative coefficients.

In order to overcome these difficulties,
we move to the grand potential $\Omega$, defined as
\begin{equation}
    Z_\xi = \exp(-\beta\Omega_\xi).
    \label{eq:Omega}
\end{equation}
$\Omega_\xi$ serves as a cumulant-generating function for correlations functions~\cite{Negele1988}
and its power series
in $\xi$ is given by:
\begin{equation}
   \Omega_\xi = \Omega_0-\frac{1}{\beta}\sum_{k=1}^\infty \xi^k \int \dd^k\VV\ D_c(\VV),
\label{eq:Omegaexp}
\end{equation}
where $\Omega_0$ is defined as
$Z_0 = \exp(-\beta\Omega_0)$.

The symbol $D_c$ indicates that unlike in
Eq.~(\ref{eq:Zexp}), the sum is to be performed over connected
Feynman diagrams only.
Using an recursion formula similar to the one introduced in Ref.~\onlinecite{Rossi2017b},
$D_c$ can be defined recursively:
\begin{equation}
    D_c(\VV) = D(\VV)-\sum_{\Ss\subsetneq\VV} \frac{|\Ss|}{|\VV|}
    D_c(\Ss) D(\VV\backslash\Ss).
    \label{eq:Dc}
\end{equation}
A derivation is given in Appendix~\ref{app:proof}. Eqs.~(\ref{eq:Dc})
and (\ref{eq:ddqmc}) allow the computation of connected diagrams as a hierarchy
of determinants at a cost of $\bigO(3^k)$.

We note that even in simple cases, the convergence radius $R$ of the series
(\ref{eq:Omegaexp}) is not infinite, with the value of $R$ depending on $h$, $U$,
and $\beta$.  Whenever $R < 1$, an order-by-order summation of the series will fail.
  We will discuss strategies to extend the convergence radius in Sec.~\ref{sec:hf}.

For convergent series ($R > 1$), one can employ the diagrammatic Monte Carlo algorithm to sample the series
(\ref{eq:Omegaexp}) by generating random vertices and computing the weight
using the recursion (\ref{eq:Dc}).  One observes that the relative statistical error
diverges exponentially with diagrammatic order $k$~\cite{Rossi2017a}, which requires truncation
of the series to a finite order $\kmax$.

\subsection{Scattering amplitude expansion}

Other than free energy, we are primarily interested in thermal correlation
function of some operators ($\hat{X}_1,\ldots,\hat{X}_m$):
\begin{equation}
  \langle \hat{X}_1 \ldots \hat{X}_m \rangle :=
  \frac 1Z \Tr[\ee^{-\beta (\hat{H} - \mu \hat{N})} \TT(\hat{X}_1 \ldots \hat{X}_m)],
\end{equation}
in particular the single-particle Green's function:
\begin{equation}
   G_{ba}(\tau) = -\langle \cop_b(\tau) \cdag_a(0)\rangle.
\end{equation}

\begin{figure*}
    \centering
    \includegraphics[width=\linewidth]{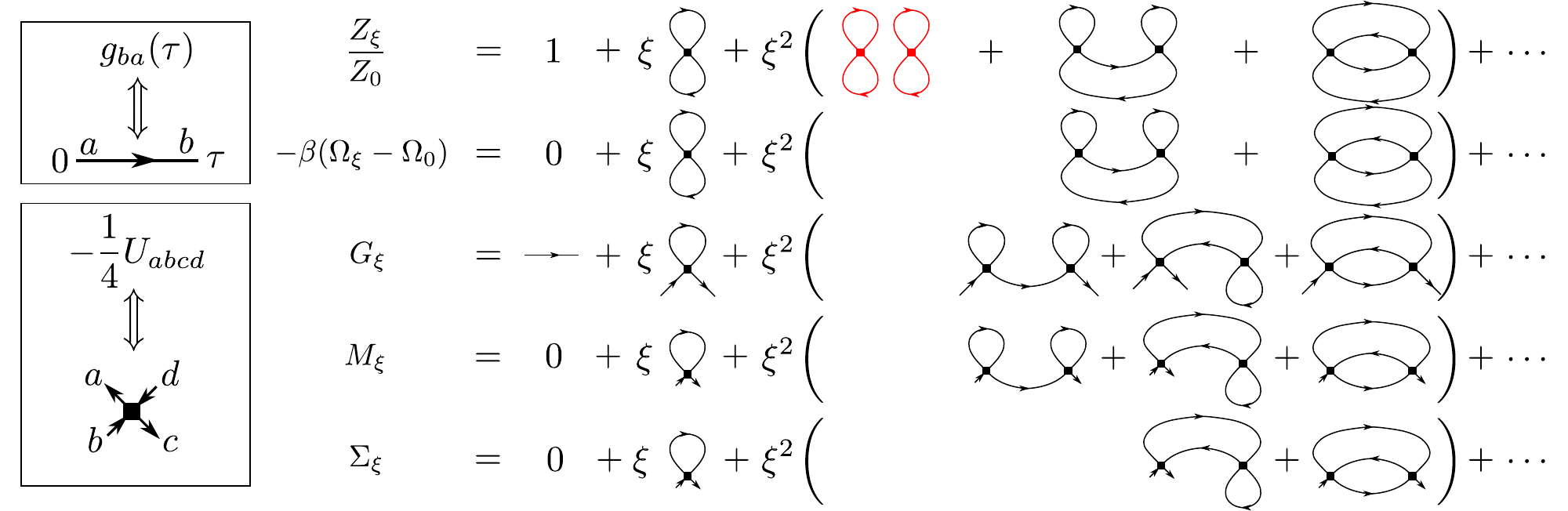}
    \caption{Schematic example of diagrams up to order 2. Diagrams shown here should be
    understood as `labeled' diagrams as described in Ref.~\cite{Negele1988}.
    Duplicate diagrams with the same topology are not shown. In the expansion of
    $Z_\xi$, the red diagram is an example of disconnected diagram, which
    is absent in the expansion of $\Omega$ due to linked cluster theorem.
    }
    \label{fig:diagrams}
\end{figure*}

One can write down a diagrammatic expansion for the Green's function similar
to Eq.~(\ref{eq:Omegaexp}) and a corresponding recursion relation~\cite{Rossi2017b}.
We instead choose to perform the expansion for a vertex-like object.

In the case of the expansion of the free energy, the corresponding one-particle
vertex is the scattering amplitude $M$ \cite{Rubtsov2005,Gull2008}, defined as:
\begin{equation}
    G(\tau) = g(\tau)
+ \int_0^\beta \dd\tau_1 \dd\tau_2\ g(\tau-\tau_1) M(\tau_1-\tau_2) g(\tau_2),
    \label{eq:m-def}
\end{equation}
where multiplication is to be understood as matrix-matrix multiplication
in spin-orbitals.
Sampling a one-particle vertex is advantageous because
it is independent of the choice of `external legs' and thus allows measurements
of both imaginary time-dependent quantities ($G$ and $\Sigma$) and
fixed-time quantities (density, kinetic energy, etc.) in the same simulation.

$M$ arises naturally as a functional derivative of the
grand potential:
\begin{equation}
    M_{ab}(\tau) = \frac{\delta(\Omega - \Omega_0)}{\delta g_{ba}(-\tau)}.
    \label{eq:dOmegadg}
\end{equation}
We show this relation in Appendix~\ref{app:mobject}.  Eq.~(\ref{eq:dOmegadg})
expresses the fact that by removing one line from a (closed) free-energy
diagram, we get an interaction correction to the Green's function, which
is exactly what the scattering amplitude encodes.

Combining Eq.~(\ref{eq:dOmegadg}) with Eq.~(\ref{eq:Omegaexp}) yields a
series expansion for $M$:
\begin{equation}
\begin{split}
   &M_{\xi,ab}(\tau) = -\frac{1}{\beta}\sum_{k=1}^\infty \xi^k
        \int \dd^k\VV \frac{\delta D_c(\VV)}{\delta g_{ba}(-\tau)}.
   \label{eq:Mexp}
\end{split}
\end{equation}
We thus need to evaluate the functional derivative of the recursion relation (\ref{eq:Dc}).

We start with the derivative of the sum of all diagrams $D(\VV)$, where we
rely on the following identity:
\begin{equation}
   \frac{\delta\det A}{\delta A_{\alpha\beta}} = (\adj A)_{\beta\alpha} := (-1)^{\alpha+\beta} \det A_{\bar\alpha\bar\beta},
   \label{eq:adj}
\end{equation}
where $A$ is an $n\times n$ matrix, $\adj(A)$ denotes the $n\times n$ adjugate matrix of
$A$, and $A_{\bar\alpha\bar\beta}$ is the $(n-1)\times(n-1)$ submatrix of $A$ with the $\alpha$-th row and
$\beta$-th column removed.
The adjugate matrix
$\adj A$ can be computed in $\bigO(n^3)$ time.
The adjugate (or cofactor) matrix arises naturally in determinantal methods as a result of the Wick's
theorem~\cite{Rubtsov2005,Gull2008,Bertrand2019}, and is often absorbed into the inverse matrix
if the matrix $A$ is not singular.
In the context of CDet, however,
care must be taken because $A$ may be singular while $\adj A$ is still meaningful~\cite{Gunacker2015}.
We elaborate on the numerical calculation in Appendix~\ref{app:adjugate}.

% Given a diagram $\VV = (v_1, \ldots, v_k)$ where $v_i = (a_i, b_i, c_i, d_i, \tau_i)$,
Combining Eq.~(\ref{eq:ddqmc}) with Eq.~(\ref{eq:adj}), we have
\begin{equation}
\begin{split}
  &\frac{\delta D(\VV)}{\delta g_{ba}(-\tau)}
  = \prod_{i=1}^k \bigg(\!-\frac{U_{a_ib_ic_id_i}}{4}\bigg)\\
  &\quad\times\sum_{\alpha,\beta}^{2n}[\adj\Gmat(\VV)]_{\alpha\beta} \delta_{a_\alpha a} \delta_{b_\alpha b} \\
  &\quad\times[\delta(\tau_\alpha-\tau_\beta-\tau) - \delta(\tau_\alpha-\tau_\beta+\beta-\tau)]\\
  &=-\sum_{\alpha,\beta}^{2n}[\Amat(\VV)]_{\alpha\beta} \delta_{a_\alpha a} \delta_{b_\alpha b} \\
  &\quad\times[\delta(\tau_\alpha-\tau_\beta-\tau) - \delta(\tau_\alpha-\tau_\beta+\beta-\tau)]
\end{split}
\label{eq:dDdg}
\end{equation}
for $0 < \tau \leq \beta$,
where $a_\alpha$, $b_\beta$, and $\tau_{\alpha(\beta)}$ takes the same meaning as
in Eq.~(\ref{eq:Gmatelem}), and we have defined the
$2k\times 2k$ matrix
\begin{equation}
    \Amat(\VV) := -\prod_{i=1}^k \bigg(\!-\frac{U_{a_ib_ic_id_i}}{4}\bigg)\adj \Gmat(\VV),
    \label{eq:Amat}
\end{equation}
which includes all connected and disconnected
amputated diagrams in which internal legs corresponding
to $\cdag_{a_\alpha}(\tau_\alpha)$ and $c_{b_\beta}(\tau_\beta)$ are removed.

For the functional derivative of a \emph{connected} free-energy diagram (\ref{eq:dDcdg}),
the sum over all diagrams in Eq.~(\ref{eq:dDdg})
with amputated legs $\Amat(\VV)$ needs to be replaced with the sum over connected diagrams with amputated
legs $\Amat_c(\VV)$:
\begin{equation}
\begin{split}
    &\frac{\delta D_c(\VV)}{\delta g_{ba}(-\tau)}
    = -\sum_{\alpha,\beta}^{2n}\bigl\{[\Amat_c(\VV)]_{\alpha\beta} \delta_{a_\alpha a} \delta_{b_\alpha b} \\
    &\quad\times[\delta(\tau_\alpha-\tau_\beta-\tau) - \delta(\tau_\alpha-\tau_\beta+\beta-\tau)]\bigr\}
\end{split}
\label{eq:dDcdg}
\end{equation}
for $0 < \tau \leq \beta$. The expansion of $M$ (\ref{eq:Mexp}) can now be
expressed in terms of $\Amat_c$ as
\begin{equation}
\begin{split}
    &[M_{\xi}(\tau)]_{\alpha\beta} = \frac{1}{\beta}\sum_{k=1}^\infty \xi^k
        \int \dd^k\VV\sum_{\alpha,\beta}^{2n}\bigl\{[\Amat_c(\VV)]_{\alpha\beta} \delta_{a_\alpha a} \delta_{b_\alpha b} \\
        &\quad\times[\delta(\tau_\alpha-\tau_\beta-\tau) - \delta(\tau_\alpha-\tau_\beta+\beta-\tau)]\bigr\}.
    \label{eq:Mexpadj}
\end{split}
\end{equation}

\begin{figure}
    \centering
    \includegraphics[width=\linewidth]{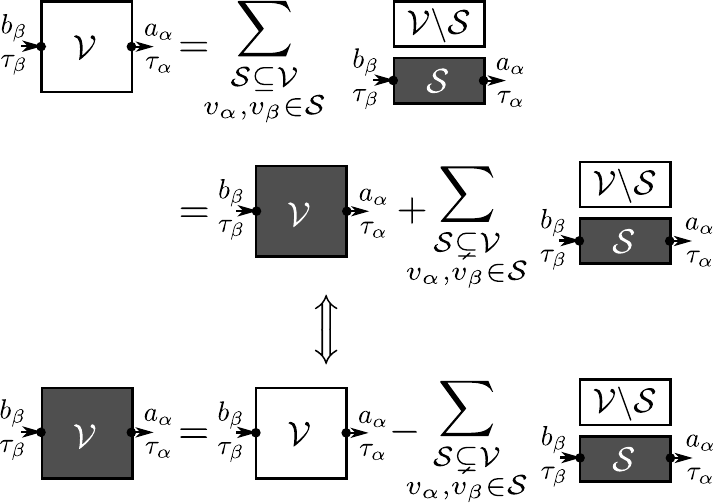}
    \caption{Schematic illustration of the recursive removal of disconnected amputated diagrams.
    Empty boxes stand for the contribution of all diagrams ($D$ or $\Amat$), and filled ones for that
    connected diagrams only ($\Amat_c$). Symbols inside boxes denote the set of vertices
    included in each component.
    The top relation shows all partitions of $[\Amat(\VV)]_{\alpha\beta}$
    into a subset fully connected to the amputated legs and a disconnected complement set.
    It is reorganized as the bottom relation which recursively defines
    $\Amat_c(\VV)$ by removing all disconnected components.
    }
    \label{fig:partition}
\end{figure}

The sum over connected amputated diagrams $\Amat_c(\VV)$ can be built up from an
recursion technique similar to Eq.~(\ref{eq:Dc}).
Defining $v_\alpha$ and $v_\beta$ as vertices where the $\alpha$-th and $\beta$-th
operators are located, respectively, diagrams in $[\Amat(\VV)]_{\alpha\beta}$ can always be
partitioned to a connected part which contains $v_\alpha$ and $v_\beta$, and the
disconnected vacuum diagrams, i.e.
\begin{equation}
    [\Amat(\VV)]_{\alpha\beta} =
    \sum_{\substack{\Ss\subseteq\VV \\ v_\alpha,v_\beta\in S}} [\Amat_c(\Ss)]_{\alpha'\beta'} D(\VV\backslash\Ss),
\end{equation}
where $\alpha',\beta'$ are row and column indices within $\Ss$ that correspond to
the row and column indices $\alpha,\beta$ in $\VV$.
Extracting the term with $\Ss=\VV$, we have the recursion relation for $\Amat_c$:
\begin{equation}
    [\Amat_c(\VV)]_{\alpha\beta} = [\Amat(\VV)]_{\alpha\beta} -
    \sum_{\substack{\Ss\subsetneq\VV \\ v_\alpha,v_\beta\in S}} [\Amat_c(\Ss)]_{\alpha'\beta'} D(\VV\backslash\Ss).
    \label{eq:m-recursion}
\end{equation}
This partitioning process is illustrated in Fig.~\ref{fig:partition}.
Since $M$ captures the interaction correction to the Green's function which
starts at the first order in interaction, the zeroth order contribution $\Amat_c(\emptyset)=0$.
For each fixed $\VV$, we apply Eq.~(\ref{eq:m-recursion}) to recursively to compute $\Amat_c(\VV)$,
which in turn yields $M$ following Eq.~(\ref{eq:dDcdg}).
Algorithmically, Eq.~(\ref{eq:m-recursion}) can be evaluated by following
Algorithm~\ref{alg:amputated}.
Algorithm~\ref{alg:amputated} runs in $\bigO(3^kk^2)$
time.

\begin{algorithm}[H]
\caption{Recursive evaluation of $\Amat_c(\VV)$\label{alg:amputated}}
\begin{algorithmic}[1]
    \Require Vertices $\VV$, $\Gmat(\VV)$ defined in Eq.~(\ref{eq:Gmat}).
    \Function{Recursion}{$\VV$, $\Gmat$}
    \If{$\VV=\emptyset$}
        \State \textbf{return} $\Amat_c(\emptyset) = 0$.
    \Else
        \State Compute $\Amat(\VV)$ from $\Gmat$ following Eq.~(\ref{eq:Amat}).
        \State Initialize $\Amat_c(\VV)\gets \Amat(\VV)$.
        \For{$\Ss\subsetneq\VV$}
            \State Compute $D(\VV\backslash\Ss)$ following Eq.~(\ref{eq:ddqmc}).
            \State $\Amat_c(\Ss)\gets$ \Call{Recursion}{$\Ss$, $\Gmat_{[\Ss,\Ss]}$}.
            \Comment{{\footnotesize $\Gmat_{[\Ss,\Ss]}$ is the submatrix of $\Gmat$ whose rows and columns correspond to the subset $\Ss$.
            Same definition applies to $[\Amat_c(\VV)]_{[\Ss,\Ss]}$.}}
            \State  Subtract $\Amat_c(\Ss) D(\VV\backslash\Ss)$ from
            $[\Amat_c(\VV)]_{[\Ss,\Ss]}$.
        \EndFor
        \State \textbf{return} $\Amat_c(\VV)$.
    \EndIf
    \EndFunction
\end{algorithmic}
\end{algorithm}

\subsection{Observables from scattering amplitude}

The electron self-energy $\Sigma$ relates
the Green's function $G$ to the non-interacting propagator $g$
via the Dyson's equation
\begin{equation}
    G(\tau)= g(\tau)
        + \int_0^\beta \dd\tau_1 \dd\tau_2 g(\tau-\tau_1)\Sigma(\tau_1-\tau_2)G(\tau_2).
    \label{eq:dyson}
\end{equation}
The expansion of the self-energy $\Sigma$ can be interpreted as
`one-particle irreducible' (1-PI) amputated diagrams, which stay connected even
when any single propagator line is removed (cf. Fig.~\ref{fig:diagrams}).
The self-energy is thus not directly sampled,
$M$ and $\Sigma$ are related to each other by \cite{Gull2008}
\begin{equation}
    \int_0^\beta \dd\tau' \Sigma(\tau-\tau')G(\tau')
     = \int_0^\beta \dd\tau' M(\tau-\tau') g(\tau')
\end{equation}
Replacing $G$ with Eq.~(\ref{eq:m-def}), we have
\begin{equation}
    \Sigma^{-1}(\iv) =  M(\iv)^{-1}+ g(\iv),
    \label{eq:sigma-from-m}
\end{equation}
where $X(\iv)$ denotes the Fourier transform of $X(\tau)$
($X=\Sigma, M, \ldots$) and $\iv$ is
a fermionic Matsubara frequency.
% \MW{Are we sure about that equation? Inverse
% of $\Sigma$ looks weird}
% \JL{This follows from $\Sigma = MgG^{-1} = Mg(g+gMg)^{-1} = (M^{-1}+g)^{-1}$}

% \paragraph{energy evaluation}
The one- and two-body contribution to the electronic energy follow from Eqs.~(\ref{eq:dyson})
and (\ref{eq:m-def}):
\begin{subequations}\label{eq:energy}
\begin{align}
    E &= E_0 + E_V\\
    E_0 &= \langle \hat{H}_0 \rangle = \frac{1}{\beta} \Tr[hG] = \sum_{ab} h_{ab}\rho_{ab}\\
    E_V &= \langle \hat{H}_V \rangle = \frac{1}{2\beta} \Tr[\Sigma G] \nonumber\\
    &= \frac{1}{2} \int_0^\beta \dd\tau \sum_{ab} M_{ab}(\tau) g_{ba}(-\tau).
\end{align}
\end{subequations}
Here $\rho_{ij}\equiv \langle \cdag_i \cop_j \rangle$ is the electron density
matrix. Note that $\hat{H}_0$ does not include the Hartree and Fock terms of the interaction.
See also Appendix~\ref{app:energy}.

\subsection{Hartree-Fock shifted Hamiltonian}\label{sec:hf}

In systems with significant electron-electron correlations where
$E_V$ has significant contribution to the full energy $E$,
the perturbation expansions in Eqs.~(\ref{eq:Omegaexp}) and (\ref{eq:Mexp})
may not converge at $\xi=1$.

In order to achieve better convergence by starting from a `better' non-interacting
solution such that $\hat{H}_0$ is closer to $\hat{H}$, we change the partition of
the Hamiltonian $\hat{H}=\hat{H}_0+\hat{H}_V$ by adding physically-motivated
counterterms to $\hat{H}_0$ and subtracting the same
terms from $\hat{H}_V$. Such an approach is
referred to the `$\alpha$-shift'~\cite{Rubtsov2005} or
as the `shifted-action'~\cite{Rossi2016} in the action formalism.

% \paragraph{counterterm}
We start by adding the simplest counterterm in the quadratic form
\begin{equation}
    \Delta \hat{H}_\shift = \sum_{ab} \shift_{ab} \cdag_a \cop_b
\end{equation}
to $\hat{H}_0$ and subtract it from $\hat{H}_V$, such that
\begin{align}
    \hat{H}_{0,\shift} &= \sum_{ab} (h_{ab} + \shift_{ab}) \cdag_a \cop_b \\
    \hat{H}_{V,\shift} &= \frac{1}{4} \sum_{abcd} U_{abcd} \cdag_a \cdag_c \cop_d \cop_b
    - \shift_{ab} \cdag_a \cop_b.
\end{align}
The total Hamiltonian $\hat{H} = \hat{H}_{0,\shift} + \hat{H}_{V,\shift}$ is unchanged,
whereas the perturbation expansion of $\hat{H}_\xi = \hat{H}_{0,\shift} + \xi \hat{H}_{V,\shift}$
can be controlled by choosing different $\shift$.
The counterterm need not be quadratic in general. Though
quadratic choices are convenient in the determinantal setup, recursion schemes have been
developed for general counterterms~\cite{Rossi2020}.

The shifted non-interacting propagator
\begin{equation}
   g^\shift(\tau) = [(-\partial_\tau + \mu)\eye - h - \shift]^{-1}_{aa'}
\end{equation}
can be seen as a Green's function with an \textit{a priori} self-energy $\shift$.

% \paragraph{Hartree-Fock counterterm}
In the molecular context, a significant contribution to electron correlations
can be obtained by the Hartree-Fock approximation.
We therefore choose $\shift$ to be the Hartree-Fock self-energy,
i.e.\ $\Sigma_\mathrm{HF}$.
%such that $g_\shift$ recovers the Hartree-Fock Green's function.
$\Sigma_\mathrm{HF}$ is given by the self-consistent equations
at finite temperature
\begin{subequations}
\begin{gather}
    [\Sigma_\mathrm{HF}]_{ab} = \sum_{cd}U_{abcd} \rho_{cd},\\
    \rho = f(h + \Sigma_\mathrm{HF} - \mu\eye).
\end{gather}
\end{subequations}
Here $f(A) = [\eye+\exp(\beta A)]^{-1}$ is the matrix-valued Fermi distribution
function, and $\mu$ is the chemical potential which may be adjusted so that
the total number of electrons in the system is adjusted to charge neutrality.

% \paragraph{diagrammatic interpretation of HF shift}

\begin{figure}
    \centering
    \includegraphics[width=\linewidth]{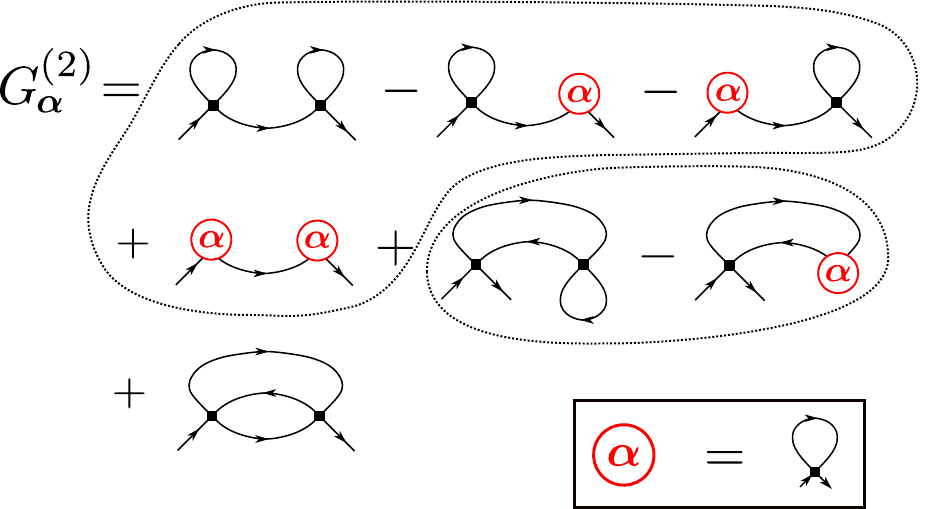}
    \caption{Schematic example of diagram cancellations due to the Hartree-Fock
    counterterm. Here we show all second-order Green's function diagrams
    generated by the counterterm, where the red circle indicates
    the counterterm $\shift$, each introduces a factor of $-1$.
    Terms in each dashed curve cancel each other, leaving only the last term.
    }
    \label{fig:shift}
\end{figure}

Diagrammatically, the Hartree-Fock shift renormalizes the propagators lines to $g_\shift$,
and an
additional effective two-point vertex $\shift$ has to be included in diagrams.
The effective vertex $\shift$ cancels any
diagram which has at least one vertex connecting to itself with \emph{exactly}
one propagator line. This removes all `tadpole' diagrams in expansions of $G$
and $M$, as well as that of $\Omega$ except for the first order diagram
whose vertex connects to itself with two propagator lines. Fig.~\ref{fig:shift}
illustrates the cancellation of such diagrams.

% \paragraph{determinantal interpretation of HF shift}
Given a specific set of vertices $\VV$, the removal of all tadpole diagrams is
achieved by replacing the $\Gmat$ matrix (\ref{eq:Gmat}) defined on internal vertices $\VV$
with:
\begin{equation}
\begin{split}
    &\Gmat(\VV) := \left[\begin{matrix}
        \bm{0}     & \bm{g}^\shift_{12}   & \cdots    & \bm{g}^\shift_{1n}\\
        \bm{g}^\shift_{21}     & \bm{0}   & \cdots    & \bm{g}^\shift_{2n}\\
        \vdots          & \vdots        & \ddots    & \vdots\\
        \bm{g}^\shift_{n1}     & \bm{g}^\shift_{n2}   & \cdots    & \bm{0}
    \end{matrix}\right]\\
    &\bm{g}_{ij}^\shift := \left[\begin{matrix}
        g^{\shift}_{b_ia_j}(\tau_i - \tau_j + 0^-) & g^{\shift}_{d_ia_j}(\tau_i - \tau_j + 0^-) \\
        g^{\shift}_{b_ic_j}(\tau_i - \tau_j + 0^-) & g^{\shift}_{d_ic_j}(\tau_i - \tau_j + 0^-)
    \end{matrix}\right].
\end{split}
\label{eq:GmatL}
\end{equation}
i.e. by setting all $2\times 2$ diagonal blocks (corresponding to self-connections
of vertices) to zero, and replacing bare propagators with $g^\shift$.
Using the modified definition of $\Gmat$ in Eqs.~(\ref{eq:ddqmc})
and (\ref{eq:Amat}), one can carry out the same recursive
calculations in Eq.~(\ref{eq:dDcdg})
to obtain corresponding connected quantities.

Note that this introduces a bias in the free-energy evaluation by setting
the first order contribution (the `dumbbell' diagram) to zero, which needs to
be corrected:
\begin{equation}
\begin{split}
\Omega_{\shift}^{(1)}
    &=\frac{1}{2} \sum_{abcd} U_{abcd} g^\shift_{ba}(0^-) g^\shift_{dc}(0^+) -
    \sum_{ab}\shift_{ab} g^{\shift}_{ba}(0^-) \\
    &= -\frac{1}{2}\sum_{ab}[\Sigma_\mathrm{HF}]_{ab}[\rho_\mathrm{HF}]_{ba}.
\end{split}
\end{equation}

In the remainder of this paper, we will always use a Hartree-Fock counterterm
and omit the $\shift$ subscripts.

\subsection{Monte Carlo integration of diagrammatic series}

% \paragraph{intro to MC integration}
Evaluations of diagrammatic series, such as Eqs.~(\ref{eq:Omegaexp}) and
(\ref{eq:Mexpadj}), can be formally summarized as
\begin{equation}
    X = \sum_{k=0}^\infty \int\dd^k\VV \CC(\VV),
\end{equation}
where $X$ is the physical variable ($G$, $M$, ...),
$\VV=(v_1,\ldots,v_k)$ denotes space time indices of internal vertices, and
$\CC$ the contribution of each fixed configuration of $\VV$ to $X$.
Here we take the `physical' value of the coupling constant $\xi=1$.
To perform a Monte Carlo integral, we introduce a cutoff $\kmax$ of the expansion
order, and an \emph{a priori} probability distribution of vertex space-time indices
$p(\VV)$ such that
\begin{equation}
    p(\VV) \geq 0,\quad\sum_{k=0}^{\kmax} \int\dd^k\VV p(\VV) \equiv 1.
\end{equation}
In addition, we require that $p(\VV)>0$ whenever $\CC(\VV)\neq 0$.
The order-$\kmax$ approximation to $X$ can be estimated stochastically as
\begin{align}
    X_{\kmax} =& \sum_{k=0}^{\kmax}\int\dd^k\VV \frac{\CC(\VV)}{p(\VV)} p(\VV)
    =\bigg\langle \frac{\CC}{p} \bigg\rangle_p\nonumber\\
    \approx& \frac{1}{\mathcal{N}}\sum_{i=1}^N \frac{\CC(\VV_i)}{p(\VV_i)},
    \quad\VV_1,\ldots,\VV_N\sim p
\end{align}
with a large number $\mathcal{N}$ of Monte Carlo samples $\{\VV_i\}$ generated following
distribution $p(\VV)$.

\begin{table*}
    \caption{\label{tab:meas}%
    Measurements for physical observables.
    % Symbols with `hats' denote quantities in Matsubara frequency representations.
    The imaginary time convolution is defined as $[f*g](\tau)=\int_0^\beta\dd\tau' f(\tau-\tau') g(\tau')$.
    $E_0^\mathrm{HF}$ and $E_V^\mathrm{HF}$ are kinetic and potential energies from
    the Hartree-Fock solution.
    }
    % \begin{ruledtabular}
    \renewcommand{\arraystretch}{1.2}
    \setlength{\tabcolsep}{10pt}
    \begin{tabular}{l|l}
        \toprule
        $X$ & $\CC(\VV)$\\
        \hline
        \hline
        $\displaystyle M_{ab}(\tau)$ &
        $\displaystyle \frac{1}{\beta}\sum_{\alpha,\beta=1}^{2|\VV|}[\Amat_c(\VV)]_{\alpha\beta} \delta_{a_\alpha a}\delta_{b_\beta b}[\delta(\tau_\alpha-\tau_\beta - \tau)-\delta(\tau_\alpha-\tau_\beta +\beta - \tau)]$\\
        \hline
        $\displaystyle{M}_{ab}(\iv_n) = \int_0^\beta\dd\tau M_{ab}(\tau)e^{\iv_n\tau}$ &
        $\displaystyle \frac{1}{\beta}\sum_{\alpha,\beta=1}^{2|\VV|}[\Amat_c(\VV)]_{\alpha\beta} \delta_{a_\alpha a}\delta_{b_\beta b}e^{\iv_n(\tau_\alpha-\tau_\beta)}$\\
        \hline
        ${G}_{ba}(\iv_n)-{g}_{ba}(\iv_n) = [{g}{M}{g}]_{ba}$ & $\displaystyle \frac{1}{\beta}\sum_{\alpha,\beta=1}^{2|\VV|}[\Amat_c(\VV)]_{\alpha\beta} g_{b a_\alpha}(\iv_n)g_{b_\beta a}(\iv_n)e^{\iv_n(\tau_\alpha-\tau_\beta)}$\\
        \hline
        $E_0 - E_0^\mathrm{HF}$ & $\displaystyle \frac{1}{\beta}\sum_{\alpha,\beta=1}^{2|\VV|}[\Amat_c(\VV)]_{\alpha\beta} \sum_{ab}h_{ab}[g_{ba_\alpha}*g_{b_\beta a}](\tau_\beta - \tau_\alpha)$\\
        \hline
        $E_V - E_V^\mathrm{HF}$ & $\displaystyle \frac{1}{2\beta}\sum_{\alpha,\beta=1}^{2|\VV|}[\Amat_c(\VV)]_{\alpha\beta} \Big\{g_{b_\beta  a_\alpha}(\tau_\beta - \tau_\alpha) + \sum_{ab}[\Sigma_\mathrm{HF}]_{ab}[g_{b a_\alpha}*g_{b_\beta a}](\tau_\beta - \tau_\alpha) \Big\}$\\
        \botrule
    \end{tabular}
    % \end{ruledtabular}
\end{table*}

% \paragraph{measurements}
Since the Green's function $G$, self-energy $\Sigma$,
as well as the total electronic energy $E=E_0+E_V$ can all be derived
from the scattering matrix $M$ using Eqs.~(\ref{eq:m-def}), (\ref{eq:sigma-from-m}),
(\ref{eq:energy}),
it is sufficient to only keep track of the amputated diagrams $\Amat_c(\VV)$
and obtain all other observables as derived quantities.
Table~\ref{tab:meas} summarizes some of these measurements.
In our implementation, we only measure the energy and ${M}(\iv_n)$ with
fermionic Matsubara frequencies $\iv_n$ on the fly, and construct
${G}(\iv_n)$ and ${\Sigma}(\iv_n)$ from ${M}(\iv_n)$ following
\begin{gather}
    {G}(\iv_n) = {g}(\iv_n) + {g}(\iv_n) {M}(\iv_n) {g}(\iv_n),\label{eq:giw-from-miw}\\
    {\Sigma}(\iv_n) - \Sigma_\mathrm{HF} = [{M}(\iv_n)^{-1} + {g}(\iv_n)]^{-1}\label{eq:sigmaiw-from-miw}
\end{gather}
for each frequency, where symbols with `hats' represent quantities in frequency
representation as matrices in spin-orbital indices.
Resampling techniques such as the jackknife or the bootstrap are applied
to avoid biased error estimations.

% \paragraph{choice of $p$}
For efficient Monte Carlo simulations, it is important to choose
the \emph{a priori} distribution $p(\VV)$ to achieve importance sampling, such that
the simulation samples more frequently when $|\CC(\VV)|$ is large and less frequently
otherwise. Since we measure multiple observables in one simulation,
we need to define such a distribution that works for all measurements.
We find in practice that the following choices provides efficient samplings
for most measurements:
\begin{align}
    p_{\Amat}(\VV) = \frac{\|\Amat_c(\VV)\|}{W_{\Amat}},&\quad W_{\Amat} = \sum_{k=0}^{\kmax}\int\dd^k\VV \|\Amat_c(\VV)\|,\\
    p_{E}(\VV) = \frac{|\epsilon(\VV)|}{W_E},&\quad W_E = \sum_{k=0}^{\kmax}\int\dd^k\VV |\epsilon(\VV)|,
\end{align}
where $\|\cdot\|$ denotes the Frobenius norm of a matrix, and
$\epsilon(\VV)$ is the energy measurement defined in Table~\ref{tab:meas}
\begin{align}
    \epsilon(\VV)=&\frac{1}{2\beta}\sum_{\alpha\beta=1}^{2|\VV|}[\Amat_c(\VV)]_{\alpha\beta}\Big\{
        g_{b_\beta a_\alpha}(\tau_\beta - \tau_\alpha) +\nonumber\\
        &+\sum_{ab}[2h+\Sigma_\mathrm{HF}]_{ab}[g_{ba_\alpha}*g_{b_\beta a}](\tau_\beta-\tau_\alpha)
    \Big\},
\end{align}
where $[f*g](\tau)=\int_0^\beta\dd\tau' f(\tau-\tau')g(\tau')$ denotes a convolution in
$\tau$. $p_E$ performs well for the energy measurements, whereas $p_\Amat$ is more
robust when measurement of $M$ is needed.

% \paragraph{normalization}
At high expansion order $\kmax$, the normalization factors $W_\Amat$ and $W_E$
are difficult to calculate analytically.
Instead, we measure an auxillary quantity whose exact value can be calculated
analytically, and normalize all other measurements against it.
For example, we can normalized against
the second-order contribution to the total energy
\begin{equation}
    E^{(2)} = \bigg\langle \frac{\epsilon(\VV)\delta_{|\VV|,2}}{p_E(\VV)} \bigg\rangle_{p_E}
    = W_E \langle \sgn[\epsilon(\VV)]\delta_{|\VV|,2} \rangle_{p_E}.
\end{equation}
Here we have chosen $p_E$ as the \emph{a priori} distribution.
Any other measurements can now be estimated as
\begin{align}
    X &= \bigg\langle \frac{\CC(\VV)}{p_E(\VV)} \bigg\rangle_{p_E}
    = W_E \bigg\langle \frac{\CC(\VV)}{\epsilon(\VV)} \bigg\rangle_{p_E}\nonumber\\
    &= E^{(2)} \frac{\langle \CC(\VV)/\epsilon(\VV)\rangle_{p_E}}{\langle \sgn[\epsilon(\VV)]\delta_{|\VV|,2} \rangle_{p_E}}.
\end{align}
Similar relations apply when we use other choices of \emph{a priori} distributions
or normalization measurements.

% \paragraph{Markov chain intro}
Once $p(\VV)$ is defined, we generate Monte Carlo samples as a
Markov chain via the Metropolis-Hastings
algorithm. From each configuration $\VV_i$, a new configuration $\VV_j$ is proposed following
some \emph{proposal probability} distribution $w^\mathrm{prop}(\VV_j|\VV_i)$. To ensure detailed
balance, an acceptance ratio $R$ is calculated after each proposal as
\begin{equation}
    R(\VV_j|\VV_i) = \frac{w^\mathrm{prop}(\VV_i|\VV_j)p(\VV_j)}{w^\mathrm{prop}(\VV_j|\VV_i)p(\VV_i)}.
\end{equation}
The proposal $\VV_i\to\VV_j$ is accepted with probability
\begin{equation}
    w^\mathrm{acc}(\VV_j|\VV_i) = \min(1, R(\VV_j|\VV_i)).
\end{equation}
This ensures the detailed balance of the Markov process, i.e.
\begin{equation}
    w(\VV_j|\VV_i)p(\VV_i)=w(\VV_i|\VV_j)p(\VV_j),
\end{equation}
where
\begin{equation}
    w(\VV_j|\VV_i) = w^\mathrm{acc}(\VV_j|\VV_i)w^\mathrm{prop}(\VV_j|\VV_i)p(\VV_i).
\end{equation}
which guarantees samples obtain the equilibrium distribution $p(\VV)$ after
thermalization.

% \paragraph{updates}

In molecular systems, due to the complexity in the multi-orbital Coulomb
interaction tensor, as well as the energy differences in non-interacting
energy levels, the configuration space of the Monte Carlo can be
uneven and may lead to ergodicity problems in the random walk.
We design the following set of updates which lead to an
ergodic random walk in the configuration space for all systems we investigate in Sec.~\ref{sec:results}.
\begin{enumerate}
    \item Vertex splitting: Split a random vertex $v=(a,b,c,d; \tau)$ to two new vertices
    $v_1=(a,b,c',d';\tau)$ and $v_2=(a',b',c,d;\tau')$. The new indices
    $a',b',c',d'$, and $\tau'$ can be proposed by some \emph{a priori} probability $p^\mathrm{ins}$.
    The proposal probability distribution for this update from order $k$ to $k+1$ is
    \begin{equation}
        w^\mathrm{prop}(v_1,v_2;k+1|v;k) = \frac{p^\mathrm{ins}(a',b',c',d',\tau')}{k}
    \end{equation}
    \item Vertex merging: Pick two random vertices $v_1=(a,b,c',d';\tau)$ and
    $v_2=(a',b',c,d;\tau')$ and merge them into $v=(a,b,c,d; \tau)$.
    The proposal probability distribution from order $k+1$ to $k$ is
    \begin{equation}
        w^\mathrm{prop}(v;k|v_1,v_2;k+1) = \frac{1}{k(k+1)}.
    \end{equation}
    \item Vertex shift in time: Update the time label $\tau$ of a vertex $v$ to a new value $\tau'$.
    \item Vertex shift in orbitals: Update one of the orbital labels $a,b,c,d$ of a vertex $v$ to a random new value.
\end{enumerate}
Vertex shift in time or orbitals are self-balancing moves, hence the acceptance
ratios shares the same form
\begin{equation}
    R(\VV_2|\VV_1) = \frac{p(\VV_2)}{p(\VV_1)}.
\end{equation}
Vertex splitting and merging are mutually inverse updates. The acceptance ratios are therefore
\begin{align}
    &R(v_1,v_2;k+1|v;k) = R(v;k|v_1,v_2;k+1)^{-1} \nonumber\\
    =&\frac{w^\mathrm{prop}(v;k|v_1,v_2;k+1)p(v_1,v_2;k+1)}{w^\mathrm{prop}(v_1,v_2;k+1|v;k)p(v;k)}\nonumber\\
    =&\frac{k+1}{p^\mathrm{ins}(a',b',c',d',\tau')}\frac{p(v_1,v_2;k+1)}{p(v;k)}.
\end{align}
There is considerable freedom in choosing $p^\mathrm{ins}$.
For all systems we study in this work, we choose $p^\mathrm{ins}$ such that
\begin{equation}
    p^\mathrm{ins}(a',b',c',d',\tau') = p^\mathrm{orb}(a',b',c',d')p^\mathrm{time}(\tau'),
\end{equation}
where $p^\mathrm{orb}(a',b',c',d')$ is uniformly distributed if the inserted indices
can form non-zero propagator connections and zero otherwise, and
\begin{equation}
    p^\mathrm{time}(\tau') = \varphi(|\tau' - \bar{\tau}_\VV|)
\end{equation}
where $\bar{\tau}_\VV = \frac{1}{k}\sum_{i=1}^k \tau_k$ is the average time coordinate
of the existing vertices, and we choose $\varphi(\tau)$
as a function in $[0,\beta]$ which has more weight
near $\tau=0$ and $\beta$ but still non-negligible weight
in between.
Since the Hartree-Fock propagators decay exponentially away from $0$ and $\beta$,
this makes sure that the new vertex are more likely to stay close to existing
vertices so that the resulting configuration has sizable contribution.
In our implementation, we define
\begin{equation}
    f(\tau) = \frac{\lambda}{\arctan( \beta\lambda)}\bigg\{
        \frac{1}{1+(\tau\lambda)^2} + \frac{1}{1+[(\beta-\tau)\lambda]^2}
    \bigg\},
\end{equation}
as a Lorentzian distribution
where $\lambda$ is an estimation of the overall energy scale of the system
proportional to e.g.\ the standard deviation of the Hartree-Fock energy levels.

\section{Results}\label{sec:results}

\subsection{Series convergence}

We first present a test of our method on a minimal molecular system: \ce{H_2} in the STO-6g
basis set~\cite{Pople1969}.
Two hydrogen atoms are placed at distance $r$ and finite temperature $T=1/\beta$. The basis set only
contains the $1\mathrm{s}$ orbital in each atom.
This setup allows us to easily perform exact diagonalization (ED) calculations of the
full molecular Hamiltonian at any temperature, such that exact benchmark results for our CDet results are available.

\begin{figure}
    \centering
    \includegraphics[width=\columnwidth]{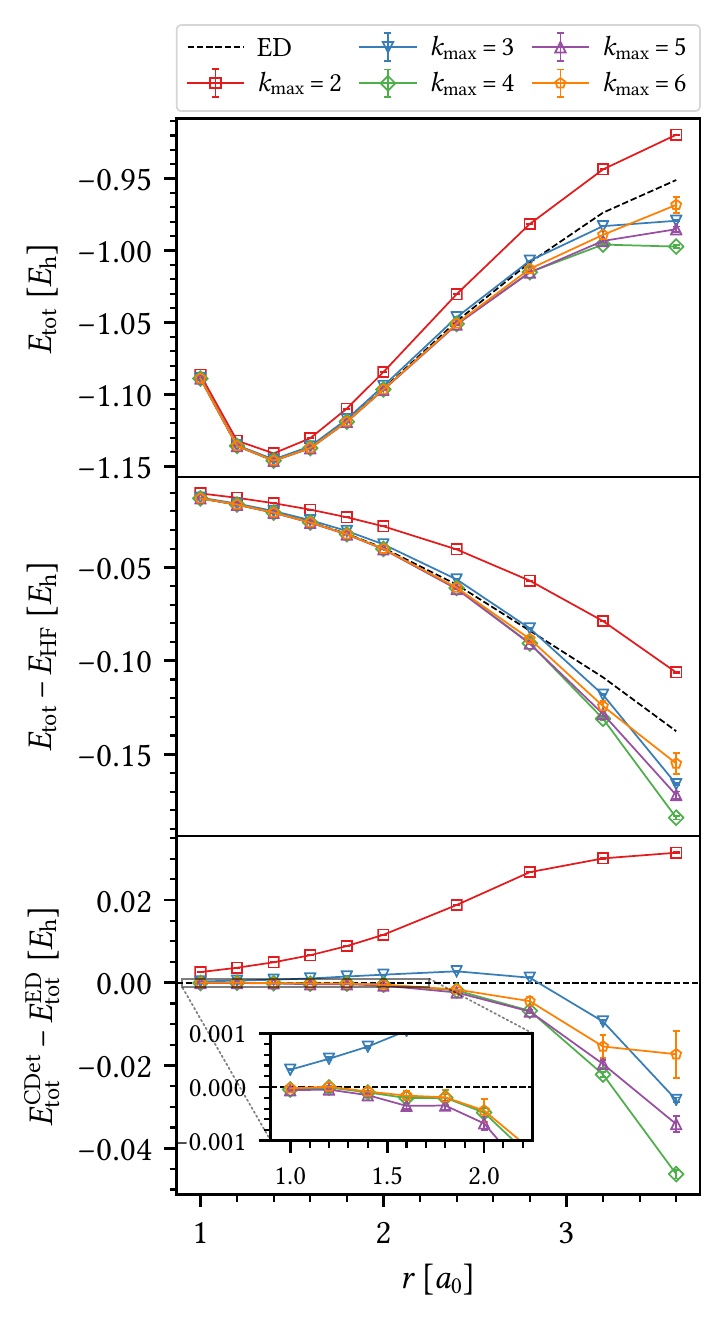}
    \caption{Total energy $E_\mathrm{tot}$ with Monte Carlo errors
    for \ce{H_2}, STO-6g, $T=50^{-1}\ \si{\hartree}$.
    Top panel: comparison of ED and CDet at different $k_{\max}$.
    Middle panel: total energy with Hartree-Fock contribution removed.
    Bottom panel: difference between ED and CDet.}
    \label{fig:h2sto-etot}
\end{figure}

In Fig.~\ref{fig:h2sto-etot}, we compare the total energy $E_\mathrm{tot}$
from CDet with
order truncation $\kmax$ up to 6 to the ED energy at $T=50^{-1}\ \si{\hartree}$,
both as a function of $r$.
Around equilibrium distance $r\approx \SI{1.4}{\bohr}$, the CDet energy converges
well to the ED solution.
The system moves to the strongly correlated regime
(i.e. a regime far from the Hartree-Fock solution),
as we `stretch' the molecule
by increasing $r$.
At $r>\SI{2.0}{\bohr}$ we
start to observe significant systematic deviation at $\kmax=6$.
Since the kinetic energy of electrons moving between two atoms is significantly
reduced as we increase $r$ but the long-range Coulomb
repulsion between electrons changes slowly,
the electron-electron interaction becomes more important at larger $r$,
and hence it is expected that the perturbation expansion becomes more difficult to converge.
This setup is standard in quantum chemistry~\cite{Szabo2012} and is similar in spirit to lattice model setups in which a metal-to-insulator transition is induced by gradually increasing an on-site interaction.

\begin{figure*}
    \centering
    \includegraphics[width=\linewidth]{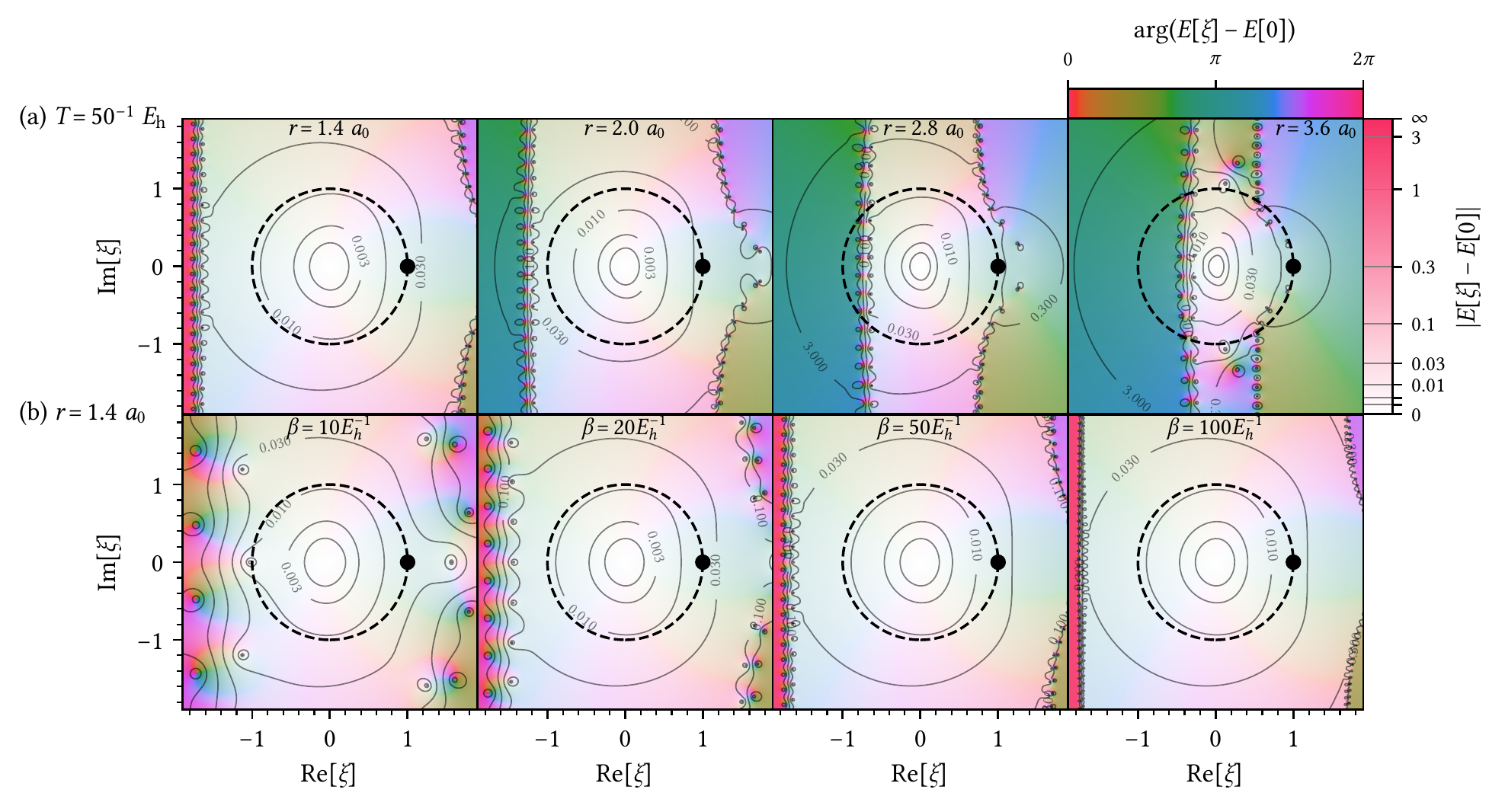}
    \caption{Analytic structure of electron total energy evaluated with ED as a function
    of the complex coupling constant $\xi$. \ce{H_2}, STO-6g.
    Colors represent complex phases and brightness indicates the magnitude (see color bars).
    The black dot at $\xi=1$ represents the `physical' result.
    The dashed black circle indicates minimum convergence radius necessary for the perturbation
    series to converge at $\xi=1$.
    (a) Effect of changing $r$ at fixed temperature $T=50^{-1}\ \si{\hartree}$.
    At $r=\SI{1.4}{\bohr}$ and $r=\SI{2.0}{\bohr}$, no singularity
    is visible in the unit circle and the series is convergent at $\xi=1$. At
    $r=\SI{2.8}{\bohr}$ and $\SI{3.6}{\bohr}$, poles appear in the unit circle, resulting in a
    divergent series at $\xi=1$.
    (b) Effect of changing $T$ at fixed $r=\SI{1.4}{\bohr}$.
    The real-axis locations of the vertical `walls' of poles
    does not change significantly as temperature decreases, while the imaginary-axis spacing
    of the poles decreases proportionally with $T$.}
    \label{fig:h2sto-poles}
\end{figure*}

Analytically, the convergence behavior is determined by the
properties of the expanded quantity (e.g.\ $E[\xi]$) as a function
of the coupling constant $\xi$ on the complex plane, similar to
the convergence analysis for many-body perturbation theory (MBPT)
calculations at $T=0$~\cite{Knowles1985,Olsen1996,Olsen1996,Hirata2015,Hirata2017,Li2019}.
We evaluate the electron energy $E[\xi]$ for complex values of $\xi$ near $\xi=0$ using ED
at $r=1.4$, $2.0$, $2.8$, and $\SI{3.6}{\bohr}$,
following
\begin{align}
    Z[\xi] =& \Tr\big\{e^{-\beta[(\hat{H}_0+\hat{H}_\shift) + \xi(\hat{H}_V-\hat{H}_\shift)-\mu \hat{N}]}\big\},\\
    E[\xi] =& \frac{1}{Z[\xi]}\Tr\big\{[(\hat{H}_0+\hat{H}_\shift/2) + \xi(\hat{H}_V-\hat{H}_\shift/2)]\nonumber\\
    &\times e^{-\beta[(\hat{H}_0+\hat{H}_\shift) + \xi(\hat{H}_V-\hat{H}_\shift)-\mu \hat{N}]}\big\},
\end{align}
where $\hat{H}_\shift$ is the Hartree-Fock counterterm introduced in Sec.~\ref{sec:hf}.
One can show via a straight forward substitution that
$E[1]$ gives the `physical' electron energy and $E[0]$ recovers the Hartree-Fock
energy.
Figure~\ref{fig:h2sto-poles}.a shows the interaction correction $E[\xi]-E[0]$
to the total energy, where the black dot represents the physical value at $\xi=1$.
Since the convergence radius of the power series around $\xi=0$ is determined by
the singularity (pole or branch cut) closest to the origin, the series is
convergent at the `physical' point $\xi=1$ if and only if there are no
singularities in the unit circle (dashed circles in Fig.~\ref{fig:h2sto-poles}).
At $r=\SI{1.4}{\bohr}$, all poles are far outside the unit circle, indicating
a rapidly convergent series.
As we increase $r$, poles move closer to the unit circle at $r=\SI{2.0}{\bohr}$,
implying a slower convergence of the series, and finally enter the unit circle
at $r=\SI{2.8}{\bohr}$ and $\SI{3.6}{\bohr}$, resulting in divergent series
at $\xi=1$.

\begin{figure}
    \centering
    \includegraphics[width=\columnwidth]{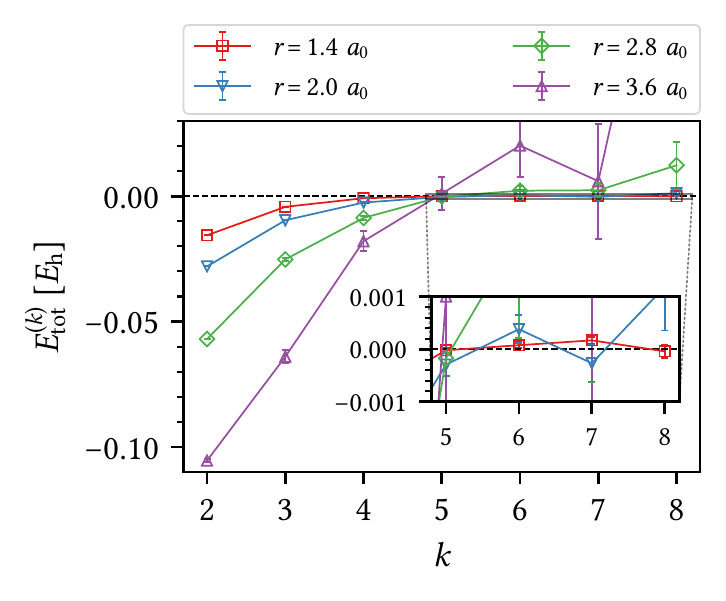}
    \caption{Contribution $E_\mathrm{tot}^{(k)}$ of each order $k$ to total energy.
    \ce{H_2}, STO-6g, $T=50^{-1}\ \si{\hartree}$. Convergence is observed at bond lengths $r \leq \SI{2.0}{\bohr}$
    but not at $r\geq \SI{2.8}{\bohr}$.}
    \label{fig:h2sto-etot-order}
\end{figure}

The analytic properties are reflected directly in
the convergence behavior of the CDet results.
For a direct comparison, we calculate the contribution
of each order $k$ to the total energy $E_\mathrm{tot}^{(k)}$ up to $\kmax=8$
for the same values of $r$,
as shown in Fig.~\ref{fig:h2sto-etot-order}.
At $r=\SI{1.4}{\bohr}$, $E_\mathrm{tot}^{(k)}$ quickly converges to zero at $k>4$.
At $r=\SI{2.0}{\bohr}$, we observe tendency to converge at $k=8$ but non-zero
systematic deviations remain.
For $r=\SI{2.8}{\bohr}$ and $\SI{3.6}{\bohr}$, no signs of convergence are
observed up to $k=8$.

\begin{figure}
    \centering
    \includegraphics[width=\columnwidth]{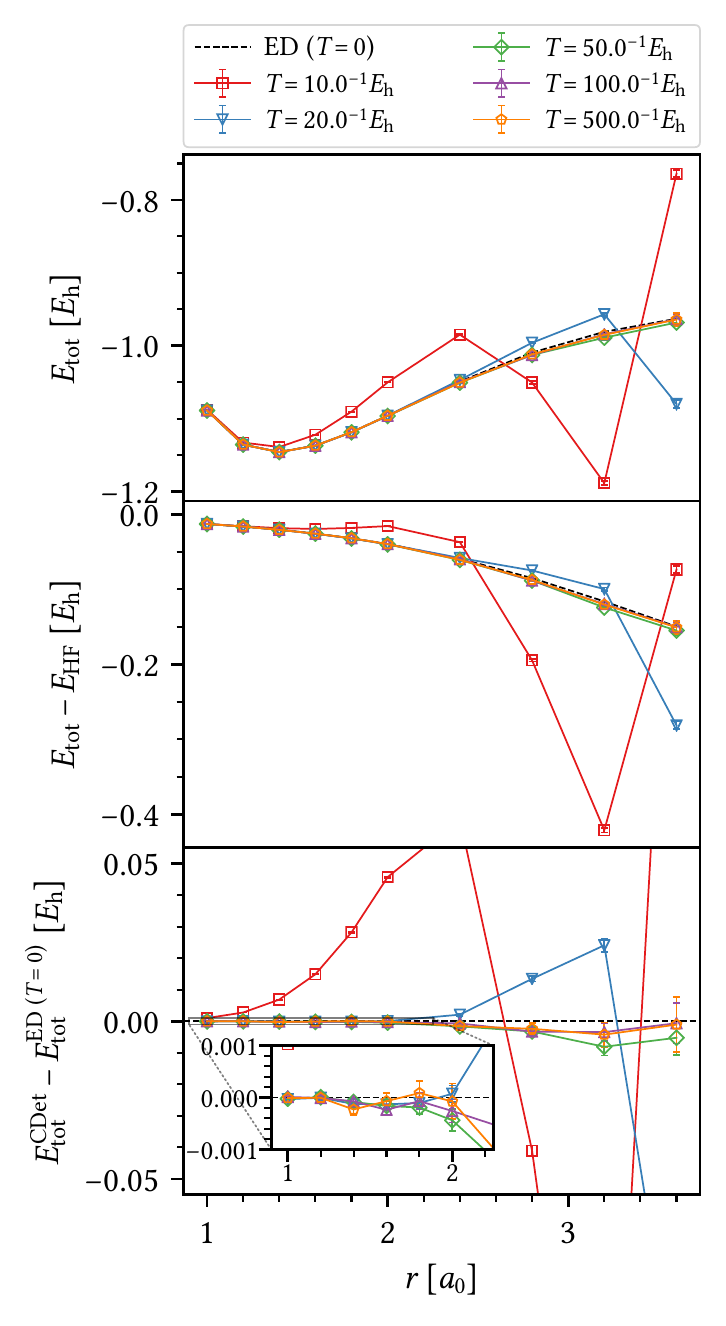}
    \caption{Temperature dependence of CDet total energy $E_\mathrm{tot}$
    for \ce{H_2}, STO-6g, $\kmax=6$.
    Top panel: comparison of ED at $T=0$ and CDet at different $T$.
    Middle panel: total energy with Hartree-Fock contribution removed.
    Bottom panel: difference between ED and CDet.}
    \label{fig:h2sto-tempdep}
\end{figure}

The CDet approach can be applied to different temperatures without adding
significant
computational cost, as we will show in Sec.~\ref{sec:cost}.
This is fundamentally different from methods such as CT-QMC, where reaching lower $T$
is only possible at an exponential cost away from half filling~\cite{Gull2011}.
In Fig.~\ref{fig:h2sto-tempdep}, we show the temperature dependence of the
CDet total energy for \ce{H_2}, STO-6g at $\kmax=6$, from
$T=10.0^{-1}\ \si{\hartree}$ down to $T=500.0^{-1}\ \si{\hartree}$, in
comparison to the ED solution at $T=0$.
All calculations use the same algorithmic setup and the same number of
Monte Carlo steps.
Convergence to the zero-temperature solution is observed as $T$ decreases,
while the stochastic error estimation does not change significantly.
Systematic deviations can be observed at similar locations ($r>\SI{2.0}{\bohr}$)
for different temperatures,
indicating similar convergence behavior for the same system at different
temperature.
This can be shown by the temperature dependence of the analytic structure of $E[\xi]$,
as plotted in Fig.~\ref{fig:h2sto-poles}.b.
As temperature is reduced, the spacing of the poles along the imaginary direction
decreases proportionally, but the real-axis locations of the vertical `walls' of
poles stay almost unchanged, which leads to similar convergence radii at different
temperature.

\begin{figure}
    \centering
    \includegraphics[width=\columnwidth]{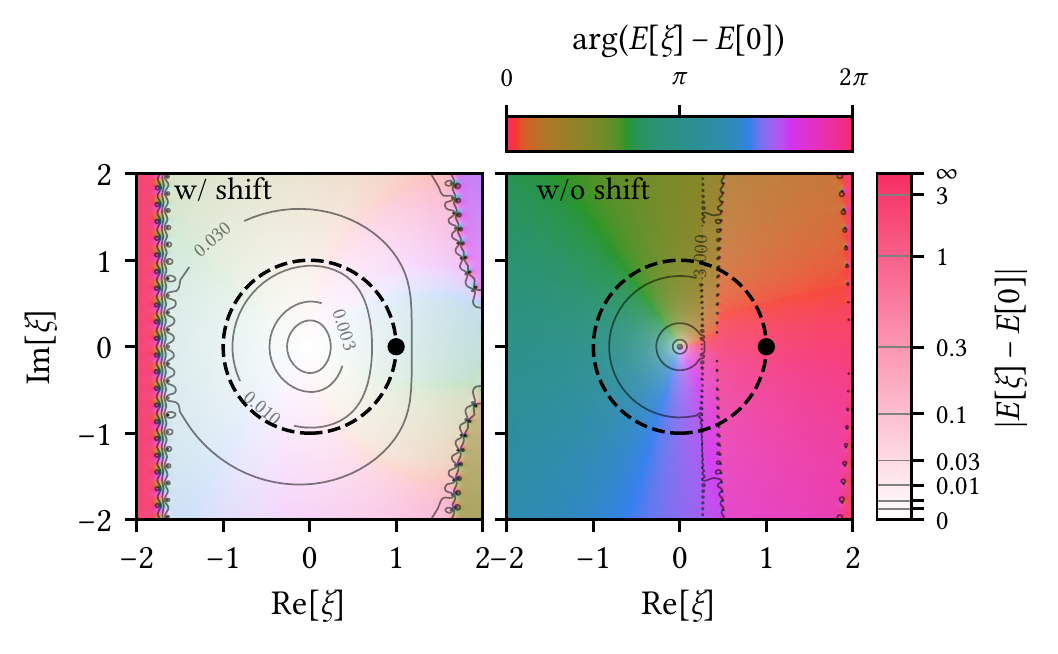}
    \caption{Analytic structure of electron total energy as a function
    of a complex coupling constant $\xi$ with and without Hartree-Fock shift.
    Left (right) panel presents values of $E[\xi]-E[0]$ evaluated on the complex plane for
    \ce{H_2}, STO-6g, $T=50^{-1}\ \si{\hartree}$ and $r=\SI{1.4}{\bohr}$ with (without) Hartree-Fock shift.}
    \label{fig:h2sto-poles-shift}
\end{figure}

The Hartree-Fock shifted action plays an important role in achieving better
series convergence in CDet. Figure~\ref{fig:h2sto-poles-shift} compares the ED
analytic structure of the total energy $E[\xi]$ with and without the Hartree-Fock
shift.
Without the shift, even for the equilibrium distance $r=\SI{1.4}{\bohr}$
(usually considered `weakly correlated'), there are poles deep inside the unit
circle, implying a highly divergent series at $\xi=1$.
In contrast, the Hartree-Fock shift pushes the poles away from the origin,
which leads to a convergent series as seen in Fig.~\ref{fig:h2sto-etot} and
Fig.~\ref{fig:h2sto-etot-order}.

\begin{figure}
    \centering
    \includegraphics[width=\columnwidth]{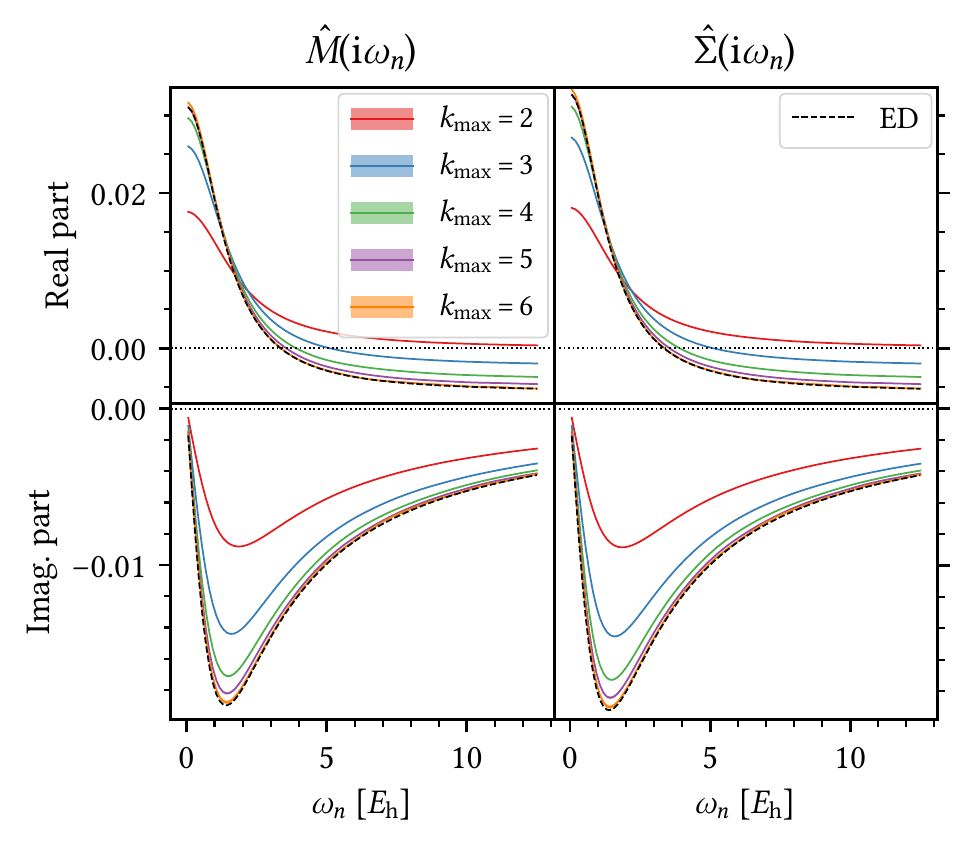}
    \caption{CDet dynamic quantities for \ce{H_2}, STO-6g, $T=50^{-1}\ \si{\hartree}$, $r=\SI{1.4}{\bohr}$.
    Shadings indicate Monte Carlo error estimates.
    Left column: Measured CDet $\mathrm{Re}{M}(\ii\omega_n)$ (top panel) and
    $\mathrm{Im}{M}(\ii\omega_n)$ (bottom panel) compared to ED.
    Right column: CDet self-energy in comparison to ED (excluding Hartree-Fock contribution $\Sigma_\mathrm{HF}$),
    top (bottom) panel showing real (imaginary) part of ${\Sigma}(\ii\omega_n)$.
    Here we show the diagonal matrix element at orbital 1 for both $M$ and $\Sigma$.}
    \label{fig:h2sto-miw-sigmaiw}
\end{figure}

\begin{figure}
    \centering
    \includegraphics[width=\columnwidth]{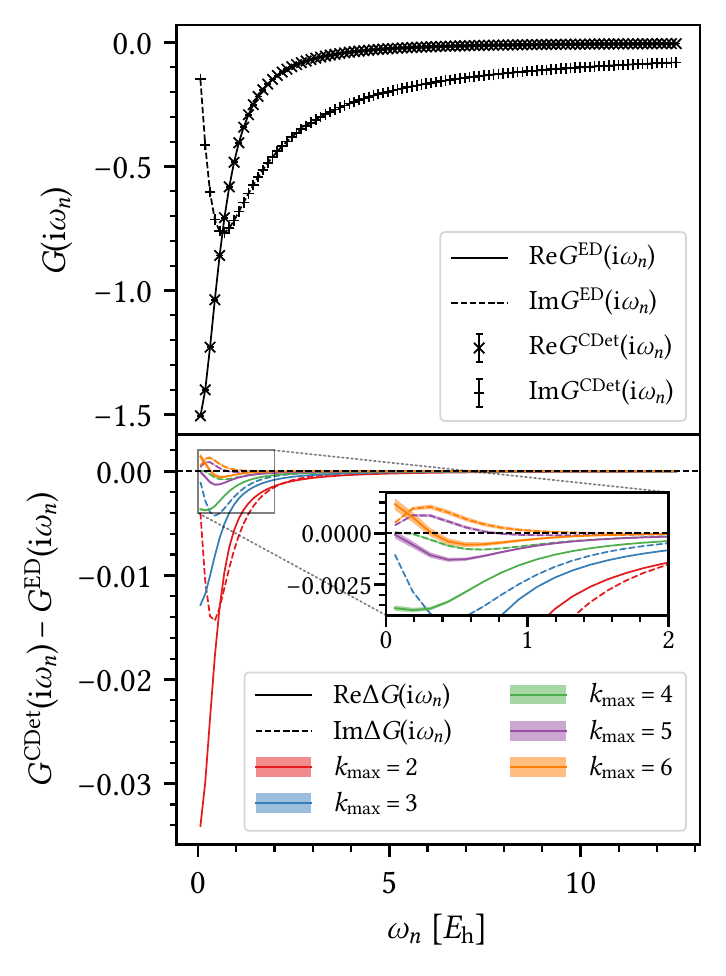}
    \caption{CDet Green's function in comparison to ED.
    \ce{H_2}, STO-6g, $T=50^{-1}\ \si{\hartree}$, $r=\SI{1.4}{\bohr}$
    Top panel: values of
    $\hat{G}(\ii\omega_n)$ at orbital 1. CDet results at $k_{\max}=6$ are plotted as symbols
    and ED values as lines. Error bars are indicated but much smaller than symbol size.
    Bottom panel: deviations of CDet results from ED at different $k_{\max}$.
    Solid (dashed) lines represent real (imaginary) part of $\hat{G}_{11}(i\omega_n)$.
    Shadings indicate stochastic uncertainties of CDet.}
    \label{fig:h2sto-giw}
\end{figure}

The CDet approach gives access to dynamic quantities, such as the Green's function $G$
and the self-energy $\Sigma$, through the scattering amplitude $M$.
The left column of Fig.~\ref{fig:h2sto-miw-sigmaiw} shows the CDet measurement of $\hat{M}(\ii\omega_n)$
in Matsubara frequency space up to $\kmax=6$
for \ce{H_2}, STO-6g at $r=\SI{1.4}{\bohr}$ and
$T=50^{-1}\ \si{\hartree}$.
As we increase the expansion order, CDet results gradually converge
to the ED solution (black lines), and at order 6 we observe only a small systematic error
due to order truncation.
The CDet self-energy $\Sigma$ is calculated from $M$ following Eq.~(\ref{eq:sigmaiw-from-miw}). Both quantities exhibit similar behavior, as shown in the right column of Fig.~\ref{fig:h2sto-miw-sigmaiw}.
At order 3 and higher, the real part of $\hat{\Sigma}(\ii\omega_n)$ takes non-zero
value at high-frequency limit, corresponding to the correction to the frequency-independent
Hartree-Fock self-energy $\Sigma_\mathrm{HF}$.
The CDet Green's function, derived from $M$ following Eq.~(\ref{eq:giw-from-miw}),
is shown in Fig.~\ref{fig:h2sto-giw}.
Good agreement with ED is observed at $\kmax=6$ on the top panel, where both
the Monte Carlo error estimation and the systematic error due to order truncation
is much smaller than the symbol size.
The bottom panel shows convergence of CDet Green's function to ED by increasing
$\kmax$, with a small but visible systematic deviation at low frequency when $\kmax=6$.

\begin{figure}
    \centering
    \includegraphics[width=\columnwidth]{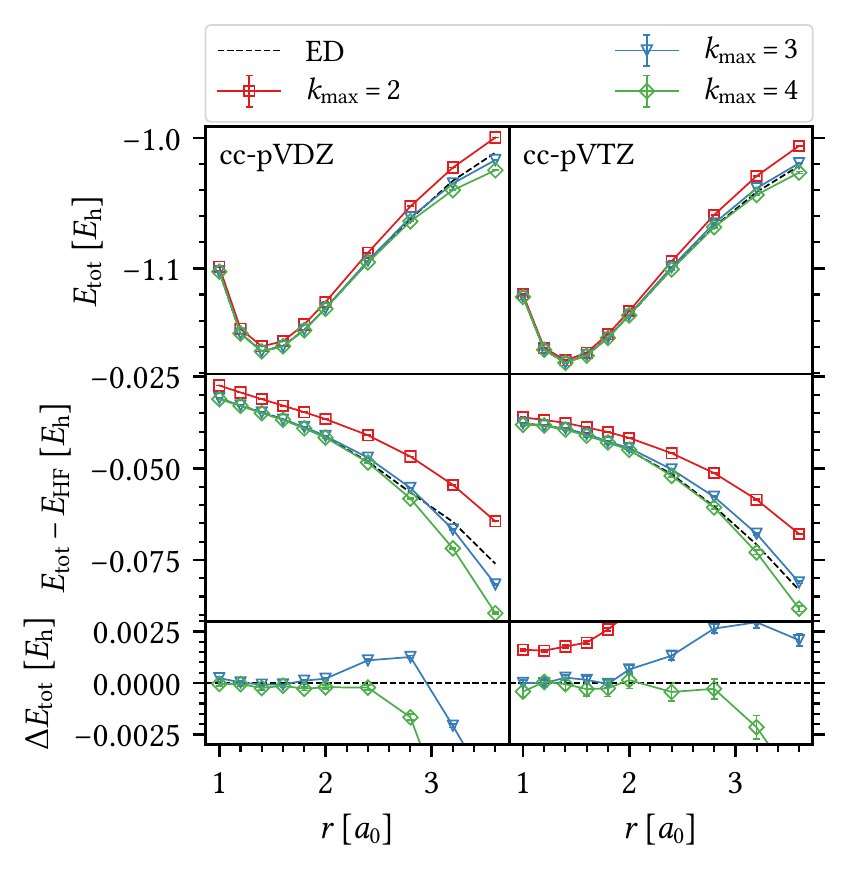}
    \caption{Total energy $E_\mathrm{tot}$ with Monte Carlo errors
    for \ce{H_2} with cc-pVDZ (left column) and cc-pVTZ(right column) basis sets,
    $T=50^{-1}\ \si{\hartree}$.
    Top panels: comparison of ED and CDet at different $k_{\max}$.
    Middle panel: total energy with Hartree-Fock contribution removed.
    Bottom panels: difference between ED and CDet.}
    \label{fig:h2vdz-h2vtz-etot}
\end{figure}

The generality of our CDet implementation allows a straightforward extension to much larger
basis sets.
Going beyond the minimal basis, we compute the CDet total energy of \ce{H_2}
using cc-pVDZ and cc-pVTZ basis sets with 10 and 28 orbitals in total, respectively,
and compare to the ED solution as shown
in Fig.~\ref{fig:h2vdz-h2vtz-etot}.
For $r < \SI{2.0}{\bohr}$, CDet gives decent convergence to ED at $\kmax=4$, with both
stochastic and systematic error below $\SI{1}{\milli\hartree}$.
The \ce{2s} and \ce{2p} orbitals added by cc-pVDZ basis and \ce{3s}, \ce{3p} and \ce{3d}
orbitals by cc-pVTZ basis are mostly unoccupied, and the most electron excitation
occur near the lowest \ce{1s} orbitals.
Consequently, the convergence behavior and computational cost of CDet do not change significantly
from the minimal basis STO-6g.

\begin{figure}
    \centering
    \includegraphics[width=\linewidth]{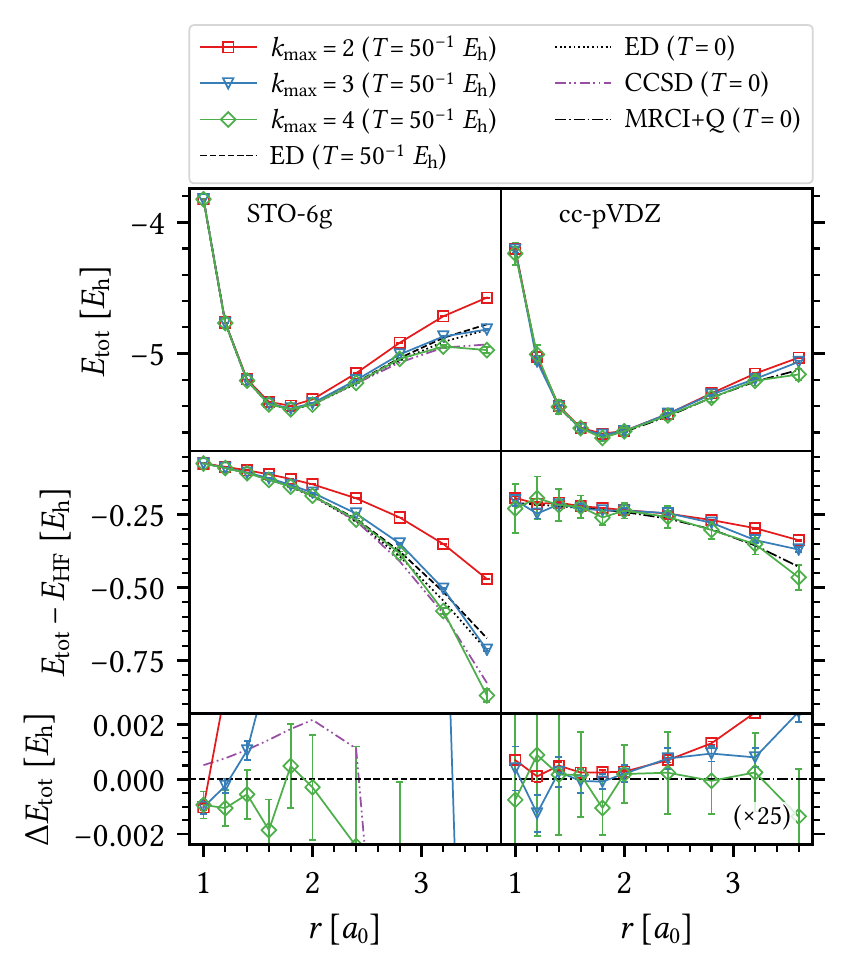}
    \caption{Total energy $E_\mathrm{tot}$ with Monte Carlo errors for \ce{H_{10}} with
    STO-6g (left column) and cc-pVDZ (right column) basis.
    ED results are used as reference for STO-6g and MRCI+Q ($T=0$) from
    Ref.~\onlinecite{Motta2017} for cc-pVDZ.
    Top panels: comparison of reference data and CDet at different $k_{\max}$ at
    finite temperature $T=50^{-1}\ \si{\hartree}$, along with ED and CCSD results at $T=0$ for STO-6g basis.
    Middle panel: total energy with Hartree-Fock contribution removed.
    Bottom panels: difference between CDet and reference data at finite temperature,
    (for STO-6g) in comparison to difference between CCSD and ED at zero temperature.}
    \label{fig:h10sto-h10vdz-etot}
\end{figure}

Finally, we extend our method to bigger molecules by adding more hydrogen atoms
to the system.
We consider a chain of 10 hydrogen atoms on a straight line with equal spacing
$r$, the same benchmark system used in Ref.~\onlinecite{Motta2017}.
At minimal basis STO-6g, all ten \ce{1s} orbitals contribute equally to the active
space of 10 electrons.
Compared to \ce{H_2} with cc-pVDZ, which has the same number of orbitals,
\ce{H_{10}} with STO-6g has more orbitals relevant to electron correlations,
and the cost of CDet is higher (for a detailed analysis see Sec.~\ref{sec:cost}).
The left column of Fig.~\ref{fig:h10sto-h10vdz-etot} plots the CDet total energy up to $\kmax=4$
in comparison to ED solution at $T=50^{-1}\ \si{\hartree}$.
Convergence within \SI{5}{\milli\hartree} is achieved at $\kmax=4$ for $r<\SI{2.4}{\bohr}$,
and systematic deviations are evident for $r>\SI{2.4}{\bohr}$.
Similar behavior can be found in the the zero-temperature coupled cluster (CCSD)
result (dotted lines), as both methods rely on the perturbative expansions
of electron-electron interactions in different forms.
The computational cost becomes much higher as we go to a bigger basis for \ce{H_{10}}.
With cc-pVDZ, there are 50 atomic orbitals in total, with potential excitations
to the empty orbitals from all 10 electrons.
As shown in the right column Fig.~\ref{fig:h10sto-h10vdz-etot}, CDet still agrees with the reference method
(MRCI+Q data from Ref.~\onlinecite{Motta2017} at $T=0$) for small values of $r$,
but the Monte Carlo errors are significantly larger.
Although our generic implementation has achieved decent extensibility without
fine tuning for each specific system,
more efficient Monte Carlo estimators and sampling schemes as well as
analytical resummation techniques should advance the limit of CDet
to more complex systems.

\subsection{Analysis of computational cost}\label{sec:cost}

\begin{figure*}
    \centering
    \includegraphics[width=\linewidth]{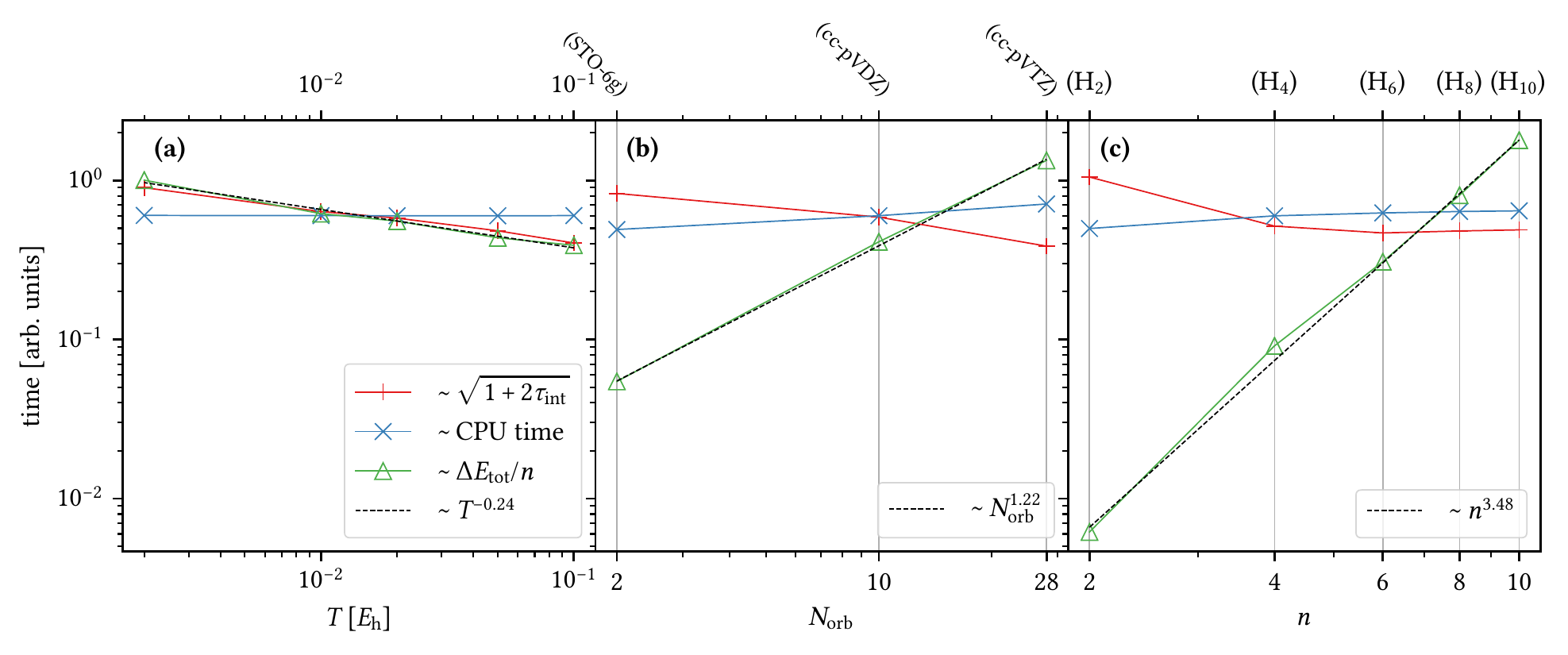}
    \caption{Empirical cost analysis of CDet simulations of hydrogen chain
    $\mathrm{H}_n$ at $r=\SI{1.4}{\bohr}$.
    In each panel, all simulations are carried out using the same setup of Monte
    Carlo updates and number of iterations.
    We estimate the contribution of integrated autocorrelation time $\tau_\mathrm{int}$
    to the stochastic error (blue), the computational cost (orange), and the total stochastic
    uncertainty of energy $\Delta E_\mathrm{tot}$ (green) for each simulation,
    and scale them to the same range on double-logarithmic plots.
    (a) Temperature dependence, \ce{H_2}, STO-6g, $\kmax=6$.
    (b) Basis set dependence, \ce{H_2}, $T=50^{-1}\ \si{\hartree}$, $\kmax=4$.
    (c) System size dependence, $\mathrm{H}_n$, cc-pVDZ, $T=50^{-1}\ \si{\hartree}$, $\kmax=4$.
    }
    \label{fig:cost}
\end{figure*}

The computational cost of a Markov chain Monte Carlo simulation, measured as the computational time needed for reaching a result for observable $X$ within a desired accuracy $\Delta X$, is determined by three factors. First, the cost of each individual update, which is $O(k^3 2^k)$ for a configuration at expansion order $k$ according to Algorithm~\ref{alg:amputated}. Second, the number of configuration updates needed to reach an independent sample by transversing a Markov chain of potentially correlated configurations, described by the integrated autocorrelation time $\tau_\mathrm{int}$. Finally, the variance $\mathrm{Var}(X)$ of the estimator of the quantity of interest (Table~\ref{tab:meas}), such that
\begin{equation}
    \Delta X = \sqrt{\frac{\mathrm{Var}(X)}{N}(2\tau_\mathrm{int}+1)}.
\end{equation}
To assess the computational cost of our CDet implementation
for reaching a certain uncertainty level, as well as how the effort changes
with respect to temperature, choice of basis set, and system size, we perform
a series of simulations of convergent series for the hydrogen chain
$\mathrm{H}_n$
with the same Monte Carlo updates and measurements for a fixed number of
Markov chain iterations.
In Fig.~\ref{fig:cost}, we show estimates of autocorrelation effects, actual computational costs,
and stochastic uncertainties in total energy, as functions of
temperature $T$, the number of orbitals $N_\mathrm{orb}$, or the number of
hydrogen atoms $n$ in log-log plots.
We rescale the $y$-values by an arbitrary factor to emphasize the respective scaling of these quantities in the same plot.

Figure~\ref{fig:cost}.a shows the temperature dependence of CDet simulations
of a fixed system (\ce{H_2}, STO-6g, $r=\SI{1.4}{\bohr}$) at $\kmax=6$.
We observe that the simulation time does not change significantly as we
decrease temperature, indicating similar distributions of the expansion order
(usually tilted to the highest order).
The error estimate in total energy follows almost the same tendency as the
factor of the autocorrelation effect $\sqrt{2\tau_\mathrm{int}+1}$,
indicating the underlying energy estimator does not have strong temperature
dependence.
The autocorrelation effect shows a slow power-law increase as temperature is
lowered, implying that our Monte Carlo updates remain efficient at low temperature.

A similar analysis is shown in Fig.~\ref{fig:cost}.b for the basis set dependence
of the same system (\ce{H_2}, $r=\SI{1.4}{\bohr}$) at fixed temperature.
We perform CDet simulations with $\kmax=4$ for basis sets STO-6g, cc-pVDZ, and
cc-pVTZ, with 2, 10, and 28 atomic orbitals, respectively.
As we add more `virtual' orbitals to the system, the computational time increases
slowly, and the autocorrelation time even decreases as the additional
orbitals improve the connectivity of Monte Carlo configurations.
However, the stochastic error shows a different trend from the autocorrelation
effect and increases (a fit with a power law results in $\sim N_\mathrm{orb}^{1.22}$),
meaning that the additional orbitals introduce more diagrammatic configurations
with alternating signs that lead to stronger Monte Carlo fluctuations.

As we increase the systems size in Fig.~\ref{fig:cost}.c by adding more hydrogen
atoms, the stochastic error (normalized by the system size $n$) at fixed computational time increases
with a much larger power law than Fig.~\ref{fig:cost}.b (fitted $\sim n^{3.48}$), while the autocorrelation time barely changes.
This implies that adding electrons that contribute to excitations near the Fermi level rapidly increases the complexity of the diagrammatics. The result is very different from the situation where additional basis states for the same number of electrons are added (Fig.~\ref{fig:cost}.b).

The behavior illustrated in Fig.~\ref{fig:cost}.c also differs  from diagrammatic Monte Carlo applications with short-range or on-site
interactions, which are formulated directly in the thermodynamic limit~\cite{Prokofev1998b,Prokofev2008a,Prokofev2008b}
and usually do not show strong scaling dependencies on system size.
We suspect the difference is caused by the long range nature of the bare Coulomb interaction,
which introduces significant non-local electronic correlations
as the system size increases.
In this case, the use of `bold' (or `screened') interactions  instead of the bare Coulomb interactions, as performed in Ref.~\onlinecite{Motta2017},
may alleviate the problem.
However, ``bold'' methods must deal with intrinsic issues of misconvergence to unphysical solutions~\cite{Kozik2015}. Moreover,
adapting such a method to the CDet framework requires further algorithmic development.
This topic is under active development~\cite{Rossi2016,Rossi2020}.

Thus, through the empirical analysis above, we have shown that for convergent
series, the computational cost of our CDet implementation is not very sensitive
to changes in temperature or basis sets, but depends strongly on the size of
the system, or more specifically, on the number of valence electrons directly
participating in electron excitations.

\subsection{Realistic impurity: SEET for NiO}

Finally, we test our CDet implementation in a general quantum impurity problem setup that includes the coupling to a non-interacting bath.
We employ the SEET framework~\cite{Zgid2017} for the antiferromagnetic compound
\ce{NiO}, which was studied by Mott \cite{Mott1949} as one of the original correlated insulators.
Following the computational setup in Ref.~\onlinecite{Iskakov2020},
we choose fcc \ce{NiO} with lattice constant $a=\SI{4.1705}{\angstrom}$
at temperature $T\sim\SI{451}{\kelvin}$
($\beta=\SI{700}{\per\hartree}$).
The unit cell is doubled along the $[111]$ direction to capture the antiferromagnetic
ordering, which contains two nickel atoms and two oxygen atoms.
We use a $4\times 4\times 4$ momentum discretization and
the gth-dzvp-molopt-sr basis set~\cite{VandeVondele2007} with gth-pbe pseudopotential~\cite{Goedecker1996}.
The Coulomb integral is decomposed using density fitting with
the def2-svp-ri auxiliary basis~\cite{Hattig2005}.
For benchmark purposes, we select the $e_g$ orbitals of both \ce{Ni} atoms in
the unit cell as the strongly correlated `impurities', which is the minimal
choice of impurities to capture correlation effects. This yields two
independent impurities each with two orbitals.
`Non-interacting' impurity propagators are generated from a converged $GW$
simulation of the complete unit cell following the SEET framework  (for details of the computational setup see Ref.~\cite{Iskakov2020}).

\begin{figure}
    \centering

    \includegraphics[width=\columnwidth]{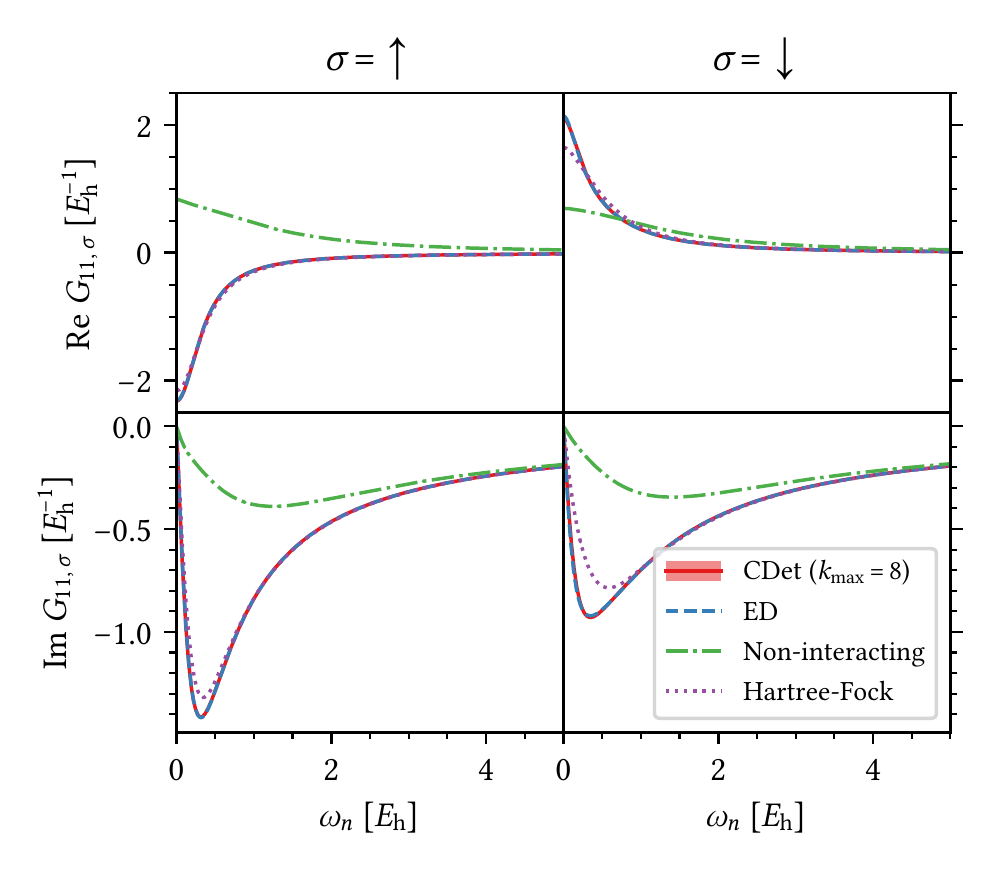}
    \caption{Matsubara Green's function for a \ce{NiO} $e_g$ impurity.
    Top (bottom) row
    shows the real (imaginary) part of the Green's functions, and
    left (right) column shows values for spin up (down).
    Red solid lines: Impurity Green's function from CDet with $\kmax=8$. Monte Carlo error
    estimations are plotted as color shadings but smaller than line width.
    Blue dashed lines: Impurity Green's function from ED, mostly overlapping
    with the CDet lines within line width.
    Green dash-dotted lines: `Non-interacting' impurity propagator with discretized
    hybridization from ED.
    Purple dotted lines: Impurity Green's function with Hartree-Fock counterterm as the starting
    point of CDet.}
    \label{fig:nio_ed}
\end{figure}

As a benchmark, we compare our CDet
impurity solver to the ED~\cite{Iskakov2018} results
used in Ref.~\onlinecite{Iskakov2020}.
ED requires the discretization of the continuous bath spectrum
and its approximation by a few states. In order to separate ED bath fitting errors from the performance of the CDet method,
we run our method for the `non-interacting' impurity Green's function $g$
corresponding to the discretized non-interacting problem solved by ED. A precise listing of all parameters and input Green's functions is given in the
supplement~\footnote{See Supplemental Material at [URL will be inserted by publisher] for
a precise listing of the input `non-interacting' Green's function $g$ (without Hartree-Fock contribution) and
the interaction tensor $V$ defined in Appendix~\ref{app:ham}.
Data is stored as a comma-separated text file with explanatory headers and comments.}.
Figure~\ref{fig:nio_ed} shows the impurity Green's functions for one
of the two $e_g$ impurities.
At $\kmax=8$, the impurity Green's function from CDet agrees with
the ED solution within line width, and the stochastic uncertainty
is almost invisible.
The spin polarization due to the antiferromagnetic ordering
is greatly enhanced in both the ED and the CDet solutions, indicating
that dynamical correlations plays an important role and are well
captured by the selected impurity.

\begin{figure}
    \centering
    \includegraphics[width=\columnwidth]{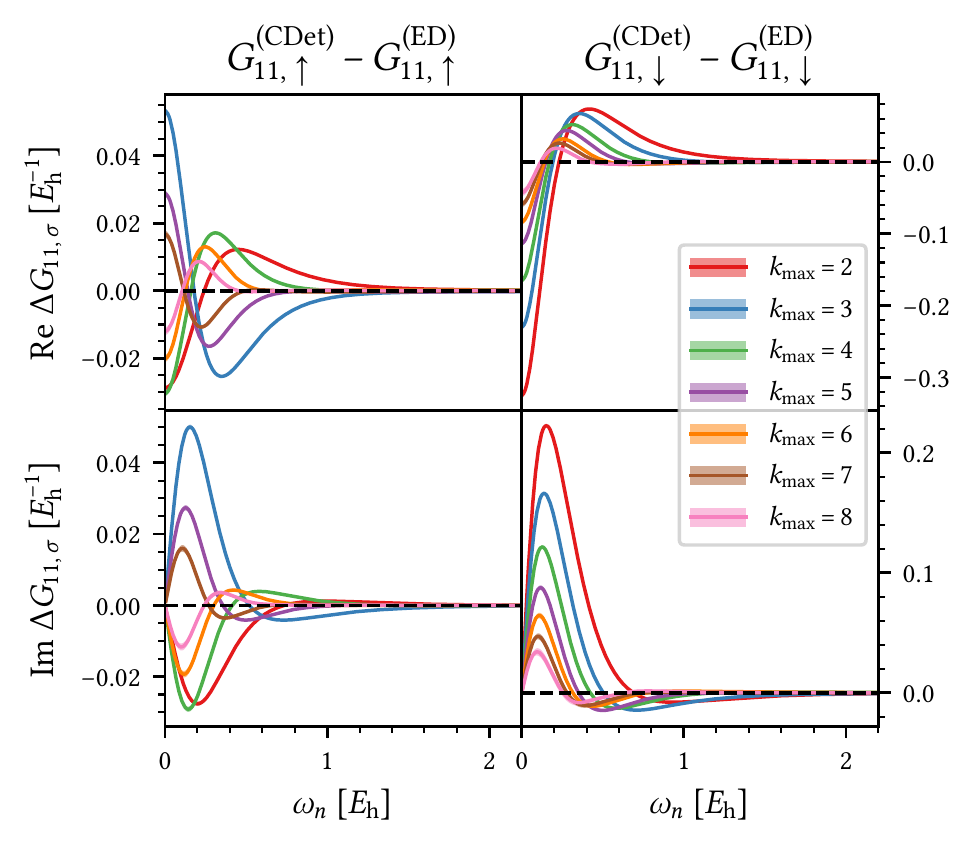}
    \caption{Difference of CDet impurity Green's functions to ED at
    different truncation order $\kmax$. Top (bottom) row shows the
    real (imaginary) part, and left (right) column shows values
    for spin up (down). Shadings indicate stochastic uncertainties of CDet.}
    \label{fig:nio_delta}
\end{figure}

In Fig.~\ref{fig:nio_delta}, we take a closer look at the convergence of CDet series
in comparison to ED by plotting the differences of CDet impurity Green's functions
to ED at different order truncations up to $\kmax=8$.
We observe that for both spins, the CDet result consistently converges to the ED
result, giving  agreement to within a percent for  $\kmax=8$.
The convergence of the spin down component is slower than spin up,
which is consistent to what can be observed in Fig.~\ref{fig:nio_ed}, i.e. the
Hartree-Fock contribution already accounts for a greater part of the overall interaction
contribution for the spin up component than for spin down.

The comparison to ED illustrates that our CDet solver can be reliably applied to
general impurity problems as part of a quantum embedding theory using
the exact same framework as developed for molecules.
We emphasize that, at the same complexity, systems with continuous bath states can be solved.
Our method is therefore a controlled method for quantum impurities with general
multi-orbital interactions and hybridizations, not limited by the systematic
error introduced by the bath discretization procedure.
The application of the solver to more complex impurities, where ED calculations are impractical,
is a topic of subsequent publications.

\section{Conclusion}\label{sec:conclusions}

In conclusion, we have presented a diagrammatic Monte Carlo method for quantum impurity models with general interactions and hybridizations
using the connected determinant formalism~\cite{Rossi2017b}.
We have tested the method at the example of molecular systems, which presents a systematic way of changing correlation strength, system size, basis size, and temperature. We have also tested our method for impurity problems occurring in  realistic quantum impurity calculations.

%In conclusion, we have have presented an application of the diagrammatic
%Monte Carlo method in the CDet formulation to molecular problems.
%We have shown examples of systems with with up to 50 orbitals which
%were tuned from weak to strong correlation (by bond stretching) and from high to
%low temperature.
%It is evident from our result that the method is not competitive with existing methods
%for quantum chemistry for the types of applications and questions tat are usually
%examined in that field.
%However, the formulation lends itself ideally to quantum impurity problem in
%DMFT and its generalizations.

Our method is formulated in the language of Green's
functions and self-energies.
As a grand-canonical finite-temperature method, it is able to describe systems
with particle number fluctuations and excited states. However, similar to other
perturbative methods, the diagrammatic series breaks down in the strong correlation
regime.
This breakdown is clearly evident in the order-by-order convergence of the
series and, as we have shown in detail, can be traced back to the pole structure
of the diagram series.

Our method fills a crucial need of impurity solvers able to treat general
four-fermion interaction and general off-diagonal hybridizations in large
multi-orbital problem.
It should therefore find applications in moderately correlated real-material
simulations such as those occurring in DMFT~\cite{Georges1996,Kotliar2006} and
SEET~\cite{Kananenka2015,Lan2015,NguyenLan2016,Zgid2017}.

Further methodological progress, such as the use of higher
order counterterms~\cite{Rossi2020}, better integration methods~\cite{Macek2020}, complex
conformal mapping techniques~\cite{Wu2017,Rossi2018b,Bertrand2019b}, and other types
of Monte Carlo updates will expand the accessible parameter regime of the method and may make simulations in the strongly correlated regime possible.

\begin{acknowledgments}
JL, MW, and EG were supported by the Simons Foundation as part of the Simons Collaboration
on the Many-Electron Problem.
During the writing phase of the paper, MW was supported by the Austrian Science
Fund (FWF) through the VeCoCo project (P30997). We thank Riccardo Rossi, Fedor
\v{S}imkovic, F\'elix Werner, and Dominika Zgid for helpful discussions.
Molecular integrals are computed using the \textsc{PySCF} library~\cite{PYSCF}.
Exact diagonalization data are computed using the \textsc{EDLib} library~\cite{Iskakov2018}.
\end{acknowledgments}

\appendix
\section{Hamiltonian}\label{app:ham}

We describe the molecular electrons using the second quantized Hamiltonian
\begin{subequations}\label{eq:ham}
\begin{align}
    \hat{H} &= \hat{H}_0 + \hat{H}_V, \\
    \hat{H}_0 &=
    \sum_{pq}\sum_{\sigma} h_{pq} \cdag_{p\sigma} \cdag_{q\sigma}, \\
    \hat{H}_V&=\frac{1}{2}\sum_{pqrs}\sum_{\sigma\sigma'} V_{pqrs}
    \cdag_{p\sigma}\cdag_{r\sigma'} c_{s\sigma'} c_{q\sigma},
\end{align}
\end{subequations}
where $\cop^{(\dagger)}_{p\sigma}$ is the electron annihilation (creation) operator
associated with orbital $\phi_p$ and spin $\sigma$.
The one- and two-body `matrix elements' are defined
as
\begin{align}
    h_{pq} &= \int \dd\mathbf{r} \phi^*_p(\mathbf{r}) \left(
        -\frac{\nabla^2}{2m_e} -
        \sum_{I}^{N_n} \frac{Z_I}{|\mathbf{r} - \mathbf{R}_I|}
    \right) \phi_q(\mathbf{r}),\\
    V_{pqrs} &= \int \dd\mathbf{r} \dd\mathbf{r}'
    \phi^*_p(\mathbf{r}) \phi_q(\mathbf{r})
    \frac{1}{|\mathbf{r} - \mathbf{r}'|}
    \phi^*_r(\mathbf{r}') \phi_s(\mathbf{r}'),
\end{align}
where $\{\mathbf{R}_I\}$ are the coordinates of the $N_n$ nuclei,
each with charge $Z_I$, and $m_e$ is the electron mass.
Eq.~(\ref{eq:ham}) defines a quantum many-body problem which
can be studied with a wide range of approximate or exact theoretical and numerical
methods.

% \paragraph{notations, antisymmetrized}
In order to combine spin and orbital indices in Eq.~(\ref{eq:ham}),
we introduce the compound notation
$\{a, b, \ldots\}$ such that
\begin{align}
    h_{ab} = h_{(p\sigma)(q\sigma')} &\equiv h_{pq}\delta_{\sigma\sigma'}\\
    V_{abcd} = V_{(p\sigma)(q\sigma')(r\lambda)(s\lambda')} &\equiv
    V_{pqrs}\delta_{\sigma\sigma'}\delta_{\lambda\lambda'}.
\end{align}
The fermionic antisymmetry can be explicitly encoded in the
anti-symmetrized interaction $U_{abcd}=V_{abcd} - V_{adcb}$,
such that Eq.~(\ref{eq:ham}) becomes
\begin{equation}
    \hat{H}_0 + \hat{H}_V = \sum_{ab} h_{ab}\cdag_a \cop_b +
    \frac{1}{4} \sum_{abcd} U_{abcd} \cdag_a \cdag_c \cop_d \cop_b.
    \label{eq:ham-anti}
\end{equation}

\section{Scattering amplitude}\label{app:mobject}

The intuition in defining the $M$ object is similar to the relation between
the self-energy $\Sigma$ and the Luttinger-Ward functional $\Phi[G]$
\begin{equation}
    \Sigma(x',x) = \frac{\delta\Phi[G]}{\delta G(x,x')},\label{eq:sigma-def-deriv}
\end{equation}
which gives the 1PI amputated diagrams with the `bold' propagator $G$~\cite{Luttinger1960}.
Here we have employed the compound space-time indices $x=(a,\tau)$.
To get the connected amputated diagrams with the `bare' propagator instead, we
define a similar relation
\begin{equation}
    M(x',x)=\beta\frac{\delta(\Omega-\Omega_0)}{\delta g(x,x')}=-\frac{\delta\log{Z/Z_0}}{\delta g(x,x')}.
    \label{eq:Mderiv-x}
\end{equation}
We show that by carrying out this functional derivative, we will recover the definition of $M$ as in Eq.~(\ref{eq:m-def}).

Switching to the action formalism using coherent state path-integrals of Grassmann variables~\cite{Negele1988},
we rewrite the partition function as
\begin{equation}
    Z=\int\DD[\cbar,c]e^{-S[\cbar,c]},
\end{equation}
where the action is given as
\begin{equation}
\begin{split}
    &S=S_0+S_V,\\
    &S_0=-\int\dd y \dd y'\cbar(y')g^{-1}(y',y)c(y),\\
    &S_V=\frac{1}{4}\int\dd\tau\sum_{abcd} U_{abcd} \cbar_a(\tau)\cbar_c(\tau) c_d (\tau)c_b(\tau).
\end{split}
\label{eq:action}
\end{equation}
Observe that
\begin{equation}
\begin{split}
    &\frac{\delta\log Z}{\delta g(x, x')} = \frac{1}{Z}
    \frac{\delta}{\delta g(x, x')} \int\DD[\cbar,c] \\
    &\quad\times\exp\left[\int\dd y \dd y'\cbar(y')g^{-1}(y',y)c(y) - S_V\right]\\
    &= \int\dd y \dd y'\frac{\delta g^{-1}(y',y)}{\delta g(x,x')}\frac{1}{Z} \int\DD[\cbar,c]
    \cbar(y') c(y) e^{-S}\\
    &= \int\dd y \dd y' \frac{\delta g^{-1}(y',y)}{\delta g(x,x')} G(y,y').
\end{split}
\end{equation}
Using the fact that for an invertible matrix $\bm{A}$,
\begin{gather}
    (\bm{A} + \delta\bm{A})^{-1} - \bm{A}^{-1} = -\bm{A}^{-1} \delta\bm{A} \bm{A}^{-1}\\
    \frac{\delta[\bm{A}^{-1}]_{ij}}{\delta\bm{A}_{kl}} = -[\bm{A}^{-1}]_{ik}[\bm{A}^{-1}]_{lj},
\end{gather}
we have
\begin{equation}
    \frac{\delta\log Z}{\delta g(x, x')} = -\int\dd y \dd y'g^{-1}(x', y)G(y, y')g^{-1}(y',x).
\end{equation}
Similarly in the non-interacting case,
\begin{align}
    \frac{\delta\log Z_0}{\delta g(x, x')}
    &= -\int\dd y \dd y'g^{-1}(x', y)g(y, y')g^{-1}(y',x)\nonumber\\
    &= -g^{-1}(x', x).
\end{align}
Putting it all together, we have
\begin{align}
    &M(x', x) = -\frac{\delta \log (Z/Z_0)}{\delta g(x, x')}\nonumber\\
    &\quad= \int\dd y \dd y'g^{-1}(x', y)[G(y, y') - g(y, y')]g^{-1}(y',x).
\end{align}
Therefore
\begin{equation}
    G(y, y') = g(y, y') + \int\dd x\dd x' g(y, x') M(x', x) g(x, y')
\end{equation}
which is exactly the same as Eq.~(\ref{eq:m-def}).

Expanding compound indices $x,y,\ldots$, Eq.~(\ref{eq:Mderiv-x}) can be rewritten
as
\begin{equation}
    M_{ab}(\tau_1,\tau_2) = \beta \frac{\delta(\Omega - \Omega_0)}{\delta g_{ba}(\tau_2,\tau_1)}.
\end{equation}
In practice, we usually work with the time-translational invariant functions
$M(\tau)$ and $g(\tau)$ instead of their two-variable form.
To that effect, we consider for $0 < \tau \leq \beta$,
\begin{equation}
\begin{split}
    &\beta\frac{\delta(\Omega - \Omega_0)}{\delta g_{ba}(-\tau)} = \sum_{a'b'}\int_0^\beta \dd\tau_1\dd\tau_2
    \frac{\beta\delta(\Omega-\Omega_0)}{\delta g_{b'a'}(\tau_2,\tau_1)}
    \frac{\delta g_{b'a'}(\tau_2,\tau_1)}{\delta g_{ba}(-\tau)}\\
    &=\sum_{a'b'}\int_0^\beta \dd\tau_1\dd\tau_2 M_{a'b'}(\tau_1-\tau_2)
    \frac{\delta g_{b'a'}(\tau_2,\tau_1)}{\delta g_{ba}(-\tau)},
\end{split}
\end{equation}
in which
\begin{equation}
\begin{split}
    &\frac{\delta g_{b'a'}(\tau_2,\tau_1)}{\delta g_{ba}(-\tau)} =
    \frac{\delta g_{b'a'}(\tau_2-\tau_1)}{\delta g_{ba}(-\tau)}\\
    &=\delta_{b'b}\delta_{a'a}[\delta(\tau_2-\tau_1+\tau)\Theta(\tau_1-\tau_2)\\
    &\quad -\delta(\tau_2-\tau_1+\tau-\beta)\Theta(\tau_2-\tau_1)].
\end{split}
\end{equation}
Therefore
\begin{equation}
\begin{split}
    &\beta\frac{\delta(\Omega - \Omega_0)}{\delta g_{ba}(-\tau)}
    =\sum_{a'b'}\int_0^\beta \dd\tau_1\dd\tau_2 M_{a'b'}(\tau_1-\tau_2)\\
    &\quad\times\delta_{b'b}\delta_{a'a}[\delta(\tau_2-\tau_1+\tau)\Theta(\tau_1-\tau_2)\\
    &\qquad -\delta(\tau_2-\tau_1+\tau-\beta)\Theta(\tau_2-\tau_1)]\\
    &=\int_0^\beta \dd\tau_1\dd\tau_2 [M_{ab}(\tau)\delta(\tau_2-\tau_1+\tau)\Theta(\tau_1-\tau_2)\\
    &\qquad -M_{ab}(\tau-\beta)\delta(\tau_2-\tau_1+\tau-\beta)\Theta(\tau_2-\tau_1)]\\
    &=M_{ab}(\tau)\int_0^\beta \dd\tau_1\dd\tau_2[\delta(\tau_2-\tau_1+\tau)\Theta(\tau_1-\tau_2)\\
    &\qquad +\delta(\tau_2-\tau_1+\tau-\beta)\Theta(\tau_2-\tau_1)]\\
    &=M_{ab}(\tau)\int_0^\beta\dd\tau_1\int_{\tau_1}^{\tau_1+\beta}\dd\tau_2
    \delta(\tau_2-\tau_1+\tau-\beta)\\
    &=\beta M_{ab}(\tau),
\end{split}
\end{equation}
which gives Eq.~(\ref{eq:dOmegadg}).
\section{Thermal expectation value of the electron energy}\label{app:energy}

The one-body energy is straightforward:
\begin{equation}
    E_0 = \langle \hat H_0 \rangle = \sum_{ab} h_{ab} \langle \cdag_a \cop_b \rangle = \sum_{ab} h_{ab} \rho_{ab}.
\end{equation}
Expression for the two-body energy term can be derived in multiple ways such as
using the equation of motion or the Schwinger-Dyson equation.
Here we provide a simple
derivation following Ref.~\onlinecite{Lin2018}.
We introduce a coupling constant
$\xi$ to the action defined in Eq.~(\ref{eq:action}) such that $S_\xi = S_0 + \xi S_V$
and $\xi \to 1$ recovers the `physical'
results. Now we have
\begin{align}
    \left. \frac{\dd Z_\xi}{\dd \xi} \right|_{\xi = 1}
    &= -\left.\int\DD[\cbar,c]S_V e^{-S_0 - \xi S_V}\right|_{\xi=1} \nonumber\\
    &= -Z \left\langle \int_0^\beta \dd\tau \frac{U_{abcd}}{4}
    \cbar_a(\tau) \cbar_c(\tau) c_d(\tau) c_b(\tau) \right\rangle \nonumber\\
    &= -Z \beta\langle \hat H_V \rangle.
    \label{eq:app:dZ1}
\end{align}
Introducing a change of variables such that $c\to c/\xi^{1/4}$
and $\cbar\to\cbar/\xi^{1/4}$, then
\begin{align}
    &\DD[\cbar,c] =
    \lim_{\mathcal{N}\to\infty} \prod_{\alpha=1}^{\mathcal{N}} \dd \cbar_\alpha \dd c_\alpha \nonumber\\
    \to& \lim_{\mathcal{N}\to\infty} \xi^{+\mathcal{N}/2} \prod_{\alpha=1}^{\mathcal{N}} \dd \cbar_\alpha \dd c_\alpha \nonumber\\
    =& \lim_{\mathcal{N}\to\infty} \xi^{+\Tr[I]/2} \prod_{\alpha=1}^{\mathcal{N}} \dd \cbar_\alpha \dd c_\alpha \nonumber\\
    =& \xi^{+\Tr[I]/2} \DD[\cbar,c]
\end{align}
where the indices $\alpha$ denote states at each discretized time point on the integration
path, $\Tr[I] = \int\dd x \delta(x, x)$ in which $x$ is the compound spacetime index,
and the plus sign on the exponent is due to the nature
of Grassmann integrals. The partition function is unaffected by the change of
variables, which now takes the form
\begin{equation}
    Z_\xi = \xi^{\Tr[I]/2}\int\DD[\cbar,c] e^{-\xi^{-1/2}S_0 - S_V}.
\end{equation}
Therefore
\begin{align}
    \left. \frac{\dd Z_\xi}{\dd \xi}\right|_{\xi = 1}
    &= \frac{\Tr[I]}{2}Z + \left.\int\DD[\cbar,c] \frac{\xi^{-3/2}}{2} S_0 e^{-\xi^{-1/2}S_0 - S_V}\right|_{\xi=1} \nonumber\\
    &= \frac{\Tr[I]}{2}Z - \frac{1}{2}\int\DD[\cbar,c] g^{-1}(x',x) \cbar(x') c(x) e^{-S} \nonumber\\
    &= \frac{\Tr[I]}{2}Z - \frac{Z}{2} g^{-1}(x',x) G(x,x')\nonumber\\
    &= \frac{Z}{2} \Tr[I - g^{-1}G]
    \label{eq:app:dZ2}
\end{align}
Comparing (\ref{eq:app:dZ1}) and (\ref{eq:app:dZ2}), we have
\begin{align}
    \langle \hat H_V \rangle &= \frac{1}{2\beta} \Tr[g^{-1}G - I]
    = \frac{1}{2\beta} \Tr[(g^{-1} - G^{-1})G]\nonumber\\
    &= \frac{1}{2\beta} \Tr[\Sigma G]
\end{align}

\section{Recursion relations for the grand potential}\label{app:proof}

The expansion of the grand potential, Eq.~(\ref{eq:Omegaexp}),
consists of connected vacuum diagrams as shown in Fig.~\ref{fig:diagrams}.
As a result of Wick's theorem (\ref{eq:ddqmc}),
all vacuum diagrams $D(\VV)$
for a fixed vertex configuration $\VV$ can be partitioned
into a connected subdiagram and the remainder of the vacuum components.
Since no external legs exist to serve as reference points for defining connectivity,
we start by picking a specific vertex $v\in\VV$ as the `reference' and consider
connectivity with respect to $v$, i.e.
\begin{equation}
    D(\VV) = \sum_{\substack{\Ss\subseteq\VV\\\Ss\ni v}}
    D_c(\Ss) D(\VV\backslash\Ss).
\end{equation}
As the choice of $v$ is arbitrary, it can be any of the $k=|\VV|$ vertices in
$\VV$, therefore
\begin{equation}
    D(\VV) = \frac{1}{|\VV|}\sum_{v\in\VV}
    \sum_{\substack{\Ss\subseteq\VV\\\Ss\ni v}}
    D_c(\Ss) D(\VV\backslash\Ss).
\end{equation}
This is equivalent to iterating all possible subsets $\Ss$ of $\VV$
where the reference $v$ can be any vertex in $\Ss$:
\begin{align}
    D(\VV) &= \frac{1}{|\VV|}\sum_{\Ss\subseteq\VV}\sum_{v\in\Ss}
    D_c(\Ss) D(\VV\backslash\Ss) \nonumber\\
    &= \sum_{\Ss\subseteq\VV}\frac{|\Ss|}{|\VV|}
    D_c(\Ss) D(\VV\backslash\Ss).
\end{align}
We now extract the term where $\Ss=\VV$ from the right hand side
and obtain the recursive formula for $D_c(\VV)$:
\begin{equation}
    D_c(\VV) = D(\VV)-\sum_{\Ss\subsetneq\VV}
    \frac{|\Ss|}{|\VV|}
    D_c(\Ss) D(\VV\backslash\Ss).
\end{equation}
The initial condition is the zeroth order contribution
$D_c(\emptyset)=0$.

A more general framework of deriving the recursion relations
using idempotent polynomials is
described in Ref.~\onlinecite{Rossi2018}. This framework does not
resort to topological arguments.
\section{Numerical computation of the adjugate matrix}\label{app:adjugate}

We calculate the adjugate $\adj(A)$ of a matrix
$A\in\mathbb{R}^{n\times n}$ numerically by first performing a rank-revealing
factorization on the matrix~\cite{Stewart1998}, such as the pivoted QR via the Householder
algorithm
\begin{equation}
    A = QDRP
\end{equation}
where $Q$ is an orthogonal matrix of Householder reflections,
$D$ is a diagonal matrix,
$R$ is an upper triangular matrix in which all diagonal elements equal 1, and
$P$ is a permutation matrix of the columns.
The rank of the matrix $r=\mathrm{rank}(A)$ is determined by
the number of nonzero diagonal elements of $D$.

If $A$ is not singular, i.e.\ $r=n$, then the adjugate is given by
\begin{align}
    \adj(A)&=\det(A)A^{-1}\nonumber\\
    &=\det(P)\det(D)\det(Q)P^T R^{-1} D^{-1} Q^T\nonumber\\
    &=\bigg[(-1)^{n_P n_Q} \prod_{i=1}^{n} d_i \bigg]P^T R^{-1} D^{-1} Q^T,
    \label{eq:app:adj-fullrank}
\end{align}
where $n_P$ is the number of transpositions in the permutation $P$,
$n_Q$ is the number of Householder reflections in $Q$,
and $d_i$ are diagonal elements of $D$.
The scaling as a function of $n$ for the complexity of calculating the adjugate is the same
as the one for calculating $A^{-1}$,
and we obtain $\det(A)$ at the same time.

If $A$ is singular, i.e.\ $r<n$, $\det(A)$ becomes zero, and
Eq.~(\ref{eq:app:adj-fullrank}) is replaced by
\begin{equation}
    \adj(A) = \bigg[(-1)^{n_P n_Q} \prod_{i=1}^{n} d_i \bigg]P^T R^{-1} \adj(D) Q^T.
\end{equation}
If $r = n-1$, there is one zero in the diagonal of $D$. Assuming $d_n=0$
and $d_i\neq 0$ for $i=1,\ldots,n-1$, the adjugate of $D$ follows directly from the
definition (\ref{eq:adj})
\begin{equation}
    \adj(D)=\mathrm{diag}\bigg(\Big[0,\ldots,0,\prod_{i=1}^{n-1}d_i\Big]\bigg).
\end{equation}
If $r < n-1$, $\adj(D)=0$, therefore $\adj(A)=0$.

In the presence of off-diagonal propagators, it is possible that the
amputated diagrams $\Amat(\VV)$ is nonzero while the vacuum diagrams $D(\VV,\emptyset)$
vanish.
Therefore it is crucial to implement the adjugate of singular matrices as
discussed above.

\end{document}